\documentclass[aip,preprint,amsmath]{revtex4-1}

\usepackage{svg}
\usepackage{float}
\usepackage[hangindent=2em, labelfont=bf, font=small]{caption}
\usepackage{subcaption}
\usepackage{xspace} % just for CDs definition of \xxx in user_commands.tex
\usepackage{cleveref}
\newcommand\xvert{0.00} % for vertival space between subfigures
\newcommand{\unit}[1]{\mathrm{\, #1}}
\newcommand{\vpre}{V_\mathrm{pre}}
\newcommand{\vcol}{V_\mathrm{col}}
\newcommand{\ntot}{n_\mathrm{tot}}

\newcommand{\ncol}{n_\mathrm{col}}

\newcommand{\td}{t_\mathrm{d}}

\newcommand{\mux}[1]{\mu_\mathrm{#1}}

\draft % marks overfull lines with a black rule on the right

\begin{document}

\title{The Physical Meaning of Time-Delayed Collection Field Transients from Disordered Devices}

\author{Markus Hu{\ss}ner}
\affiliation{Department of Engineering, Durham University, Lower Mount Joy, South Road, Durham, DH1 3LE, UK}

\author{Carsten Deibel}
\email[]{deibel@physik.tu-chemnitz.de}

\affiliation{Technische Universität Chemnitz, Straße der Nationen 62, D-09111 Chemnitz, Germany}

\author{Roderick C. I. MacKenzie}
\email[]{roderick.mackenzie@durham.ac.uk}

\affiliation{Department of Engineering, Durham University, Lower Mount Joy, South Road, Durham, DH1 3LE, UK}

\date{\today}

\begin{abstract}
Charge carrier mobility and recombination determine the performance of many opto-electronic devices such as solar cells, sensors and light-emitting diodes. Understanding how these parameters change as a function of material choice, charge carrier density and device geometry is essential for developing the next generation of devices. The Time-Delayed-Collection-Field technique (TDCF) is becoming a widely used method to measure both recombination and carrier transport with values derived from this method being widely reported for many material systems. However, most novel materials are highly disordered with a high density of trap states and standard TDCF-theory neglects the influence of these states. In this work we examine how reliable TDCF can be as a measurement technique when the device contains significant energetic disorder. We identify regimes where the results can be relied upon and where the results should be taken with more caution. Finally, we provide simple and easy to use experimental tests to help the experimentalist decide if the physical processes are dominated by trap states.
\end{abstract}


\maketitle %\maketitle must follow title, authors, abstract and \pacs

Recently much effort has been dedicated to developing electronic devices based on conducting polymers and small molecules. Many classes of devices using these material systems that have already been demonstrated include, organic photovoltaic (OPV) devices \cite{zhu2022single}, organic light emitting diodes (OLEDs)
\cite{ma2023carbene} and optical sensors\cite{202209906}. The materials offer mechanical flexibility \cite{LEE2021stretchableOptoelectronics}, and the ability to absorb and emit light over a wide range of wavelengths \cite{Ostroverkhova2016optoelectronicMaterials, YU2019crystalengineering}. 
Two key material parameters that determine device performance are charge carrier mobility and recombination rate. Charge carrier mobility describes how conductive the device is while the recombination rate determines how long carriers can survive in a device. For example in a solar cell one would ideally like to have a high charge carrier mobility and a low recombination rate to enable photogenerated carriers to exit the device before they recombine. For OLEDs one would like a high carrier mobility to minimise joule heating and a low recombination rate in all parts of the device except the emissive layer to maximise photon generation. Thus having an accurate measure of both charge carrier mobility and recombination rate is essential if materials are to be compared and evaluated in the search for more efficient devices.

However, in materials with a high number of trap states a single value of mobility is hard to define \cite{sandberg2023midgaptrap, Ostroverkhova2016optoelectronicMaterials, Haneef2020trapsorganicsemiconductors}. In general one may define an effective mobility as 
\begin{equation}\label{eq:paper_mueff}
    \mux{eff} = \frac{1}{d} \int\limits_{0}^{d}  \frac{\mux{free} n_\mathrm{free}(x)}{n_\mathrm{free}(x) + n_\mathrm{trap}(x)} \mathrm{d}x
\end{equation}
where $\mux{free}$ is the charge carrier mobility of completely free carriers, $n_\mathrm{free}$ is the density of completely free carriers and $n_\mathrm{trap}$ is the density of trapped carriers\cite{Baranovskii2018,Goehler2018}. Free carriers will reside above the mobility edge and have more energy than trapped carriers which will generally reside in mid-gap states. The expression can be better understood if one looks at two extreme cases. If $n_\mathrm{trap}=0$ and there are no trapped carriers, the effective mobility will be equal to $\mux{free}$.  If on the other hand $n_\mathrm{free}=0$, then all carriers will be trapped and $\mux{free}$  will be zero. Thus one can see that the effective mobility of an organic device depends on how the carriers are distributed in energy space.  For a layer of organic semiconductor deposited on glass with no contacts that is kept in the dark, the carrier density will be low and carriers will mainly reside in deep states resulting in a low mobility. If metallic contacts are added charges will flood into the semiconductor filling the trap states and through the pauli-exclusion principle there will be more free carriers. If the sample is then illuminated the carrier density will further increase and because many of the trap states are already filled the free states will become more populated again increasing the average charge carrier mobility.  Applying a positive voltage will cause carriers to flood into the device and through the same reasoning increase the carrier density and thus mobility. Inversely, applying a negative voltage will drag carriers out of the device and lower the mobility. Consequently, it can be seen that mobility is very much dependent upon the exact conditions under which it is measured, and it can not be assigned a single value in disordered materials. In much the same way recombination is also highly carrier density dependent.

\begin{figure}
    \centering
    \graphicspath{{paper/bilder/}}
    \def\svgwidth{0.5\linewidth}
    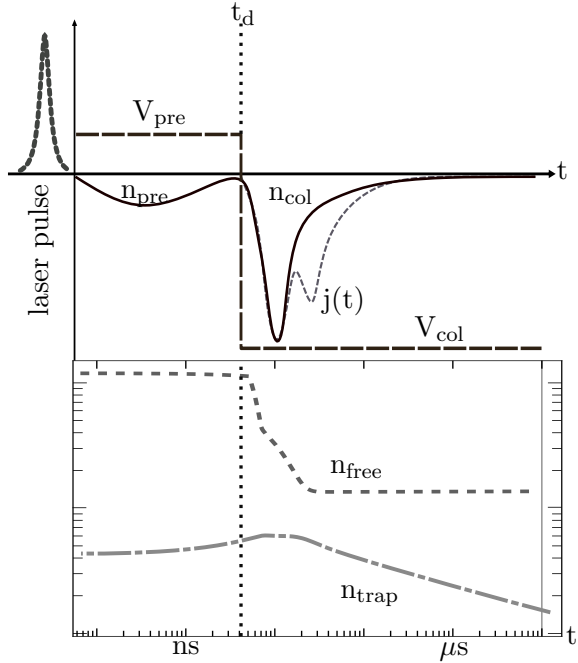
    \caption{Top: Schematic diagram of a TDCF experiment. The solid and dashed lines labelled $j(t)$ represent the current measured for a material system with symmetric and very asymmetric mobilities.  Bottom: The corresponding densities of free charge carriers $n_{\mathrm{free}}$ and trapped charge carriers $n_{\mathrm{trap}}$.}
    \label{fig:tdcf_idea_paper}
\end{figure}

The TDCF method\cite{Mort1980tdcf,Kniepert2011TDCF,popovic1983study} is often used to determine charge carrier mobility and recombination in disordered materials. The method is depicted in the top of Figure \ref{fig:tdcf_idea_paper}. The sample is held at a constant (usually positive) pre-bias $\vpre$, a short laser pulse is then applied and after the delay-time $\td$ a large negative voltage $\vcol$ is applied to extract photogenerated charge carriers that have not recombined.  By studying how the total extracted charge changes as a function of $\td$ one can obtain recombination rates, and by examining the gradient of the current transient one can measure the charge carrier mobility.

However, if one considers the description of the TDCF method in combination with the discussion above about effective carrier mobility, it can immediately be seen that as soon as the TDCF voltage pulse is applied it will start extracting carriers and thus change the mobility/energetic distribution of carriers within the device. As charge trapping is not considered in TDCF theory when attempting to recover mobility \cite{Kniepert2015diss} it is not clear what value of mobility will be extracted. Furthermore, this changing mobility would be expected to influence the charge extraction efficiency of TDCF and therefore change the measured recombination rate.  Therefore in this work we examine the validity of TDCF to measure mobility and recombination.

\pagebreak
\section{Methods}
\subsection{The model}
The TDCF method can be used to investigate a wide range of opto-electronic devices. However, in this work we focus on organic solar cells due to the prominence of the technique in the field. We study a state of the art Glass/ITO/SnO\textsubscript{2}/PM6:Y6/MoO\textsubscript{3}/Ag device structure\cite{Woepke2022transport} with an active layer thickness of 100 nm.  We chose a PM6:Y6 device because with the emergence of small molecule acceptors PM6:Y6 is quickly becoming a key model material system\cite{Shoaee2023}. Simulated device parameters are set close to previously reported experimental values\cite{Woepke2022transport}, however, they are made symmetric where necessary to simplify understanding (see SI).

The numerical model used for the simulations is described in detail elsewhere \cite{Mackenzie2020ohmicspacecharge, Mackenzie2012extractingmicroscopic}. However, 
in summary, the electric field profile within the device is calculated using Poisson's equation in one dimension. The movement of free charge carriers is described by solving the bi-polar drift diffusion equations. Conservation of particles is forced using the carrier conservation equations. As discussed in the introduction, it is important to consider charge carrier trapping in disordered devices.  Indeed, it should be noted that standard drift-diffusion models which do not consider trap sates are not valid for disordered systems as they will fail to reproduce the correct dependence of mobility and recombination rate as a function of voltage/carrier density. We therefore describe trap states below the LUMO and above the HOMO mobility edges as two exponential distributions of states, 
\begin{equation}
    \rho^\mathrm{e/h}(E) = N^\mathrm{e/h}\exp\left(-\frac{E}{E^\mathrm{e/h}_\mathrm{U}}\right)
\end{equation}
where $N^\mathrm{e/h}$ are the electron/hole trap densities at the LUMO and HOMO edge; $E^\mathrm{e/h}_\mathrm{U}$ are the characteristic electron/hole tail slope energies and $E$ is the energy relative to the LUMO/HOMO edge. This distribution is then broken up into 8 independent trap states of 0.1 eV in height and the full time domain Shockley-Read-Hall (SRH) equations are solved for each energetic range. The LUMO electron SRH trapping equation is written as

\begin{equation}
    \frac{\partial n_\mathrm{t}}{\partial t} = r_\mathrm{ec} - r_\mathrm{ee} - r_\mathrm{hc} + r_\mathrm{he}~.
\end{equation}
where $r_\mathrm{ec}$ and $r_\mathrm{ee}$ are the electron capture/escape rates  while $r_\mathrm{hc}$ and $r_\mathrm{he}$ are the hole capture/escape rates into the electron trap which describes recombination. The rates are functions of carrier depth and free/trapped carrier density\cite{Mackenzie2012extractingmicroscopic}. An analogous equation can be written for holes. Thus using this approach carrier density can be described in both energy and position space across the device.
\subsection{Mobility}

Often in OPV material systems the electron and hole mobilities are different by one or two orders of magnitude. Before looking at the influence of trap states we firstly investigate how sensitive TDCF is to asymmetric carrier mobilities in a device with no traps.

\begin{figure}
    \includegraphics[width = 0.75\linewidth]{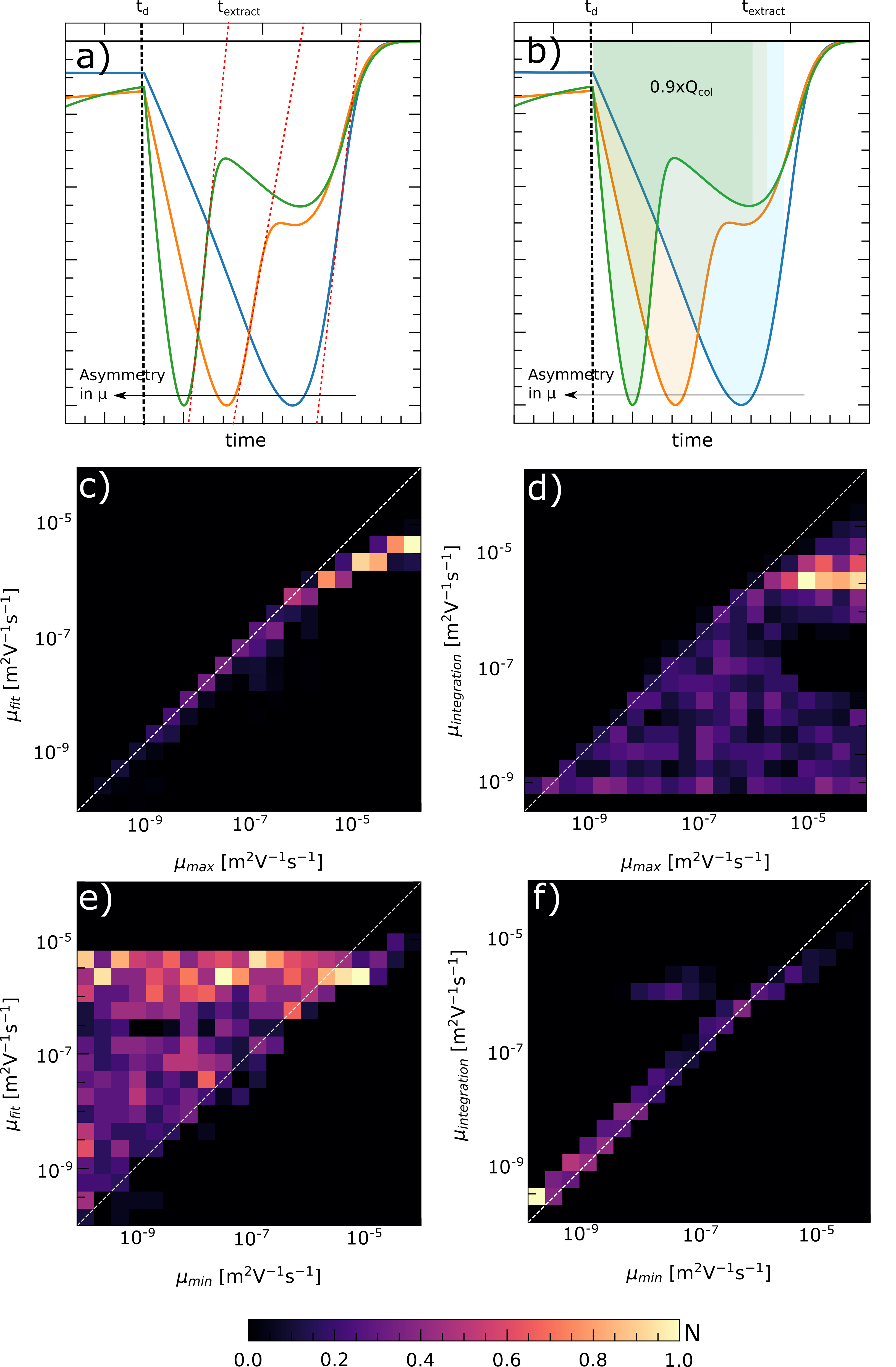}
    \caption{a) Fitting the initial linear photo-current decay to determine the mobility of the fastest charge carrier; b) Integrating the phototransient to a time when 90\% of the total charge carriers has been extracted as a way to determine the mobility of the slowest carrier; c,e) The error matrices for asymmetric input mobilities when trying to determine $\mu_{min},\mu_{max}$ using the fitting method. The fitting method can determine the mobility of the fastest charge carrier; d,f) The error matrices for asymmetric input mobilities when trying to determine $\mu_{min},\mu_{max}$ using the integration method can determine the mobility of the slowest charge carrier.}
    \label{fig:mobility_asym_paper}
\end{figure}

In TDCF experiments charge carrier mobility is obtained by firstly fitting the linear photo-current decay with the equation
\begin{equation}
    I(t) = \frac{Q_\mathrm{0}}{t_\mathrm{tr}}\left(1-\frac{t}{t_\mathrm{tr}}\right)~.
    \label{eq:fit_equ}
\end{equation}
to determine the extraction time $t_{tr}$, \cite{Kniepert2017EffectOT} where $Q_0$ is the initial charge in the device before the decay. Charge carrier mobility $\mu$ is then calculated using

\begin{equation}\label{eq:method_transittime}
    t_\mathrm{tr} = \frac{d^2}{\mu V_0}~,
\end{equation}
where $d$ is the device thickness and $V_{0}$ is the applied voltage relative to the built-in potential. The blue line in Figure \ref{fig:mobility_asym_paper}a represents a TDCF transient where the device has symmetric electron/hole mobilities, while the orange and green lines represent TDCF experiments with increasingly asymmetry in mobilities. For asymmetric cases a second peak in the current transient appears once the slower charge carrier specimen starts to reach the contacts of the device.

Theoretically, the mobility of the slower charge carrier specimen can be determined by fitting Equation \ref{eq:fit_equ} to the linear decay of the second peak. However, in practice even for symmetric mobilities it can be difficult to correctly determine the linear decay region for the fitting, let alone for the second peak. The reason for this is that charge carrier diffusion and dispersion in mobility often broaden the transient.

Figure \ref{fig:mobility_asym_paper}c plots the results of a series of simulations where a series of TDCF transients were simulated from devices with a range of electron and hole mobilities. The x-axis plots the maximum of the electron/hole mobilities for the device $\text{max}(\mu_e,\mu_h)$, while the y-axis plots the mobility extracted using TDCF. It can be seen that TDCF can accurately extract the fastest mobility.  At higher mobilities the predicted values start to deviate from the input mobilities, this is due to capacitive effects \cite{Kniepert2017EffectOT} (see also SI Fig.~\ref{fig:mobility_RC_paper}). For high mobilities the determined extraction time is in the same order as the RC-time of the device.  Figure \ref{fig:mobility_asym_paper}e, plots the minimum value of device mobility $\text{min}(\mu_e,\mu_h)$ plotted against the extracted TDCF mobility value. It can be seen that TDCF always overestimates the mobility of the slowest charge carrier.

In an attempt to recover the mobility of the slowest charge carrier, we take inspiration from Lorrmann et al.\cite{Lorrmann2014} and numerically integrate the extraction current transient to the time when 90\% of the total charge $Q_\mathrm{tot}$ is extracted to determine $t_\mathrm{tr}$. Although the threshold is more or less arbitrary, by examining Figure \ref{fig:mobility_asym_paper}d and f one can see that the method can indeed extract the value of the slower charge carrier. The mobilities in 

Until now we have considered a device without trap states where all carriers are free, $n_\mathrm{trap}=0$. However, in a realistic disordered device $n_{trap}$ will be typically be more than an order of magnitude higher than $n_{free}$. Furthermore, the values $n_{free}$ to $n_{trap}$ will change during a TDCF transient. This is depicted in the simulation in the bottom of Figure \ref{fig:tdcf_idea_paper}. It can be seen that after the voltage pulse is applied the number of free carriers rapidly drops as they are swept out the device, then much more gradually as the trapped carrier population decreases as thermal energy frees them, allowing them also to be swept out. This slow detrapping process has been previously described in Ref.~\cite{Hanfland2013CELIV}.

\begin{figure}
    \centering
    \includegraphics[width = 0.85\linewidth]{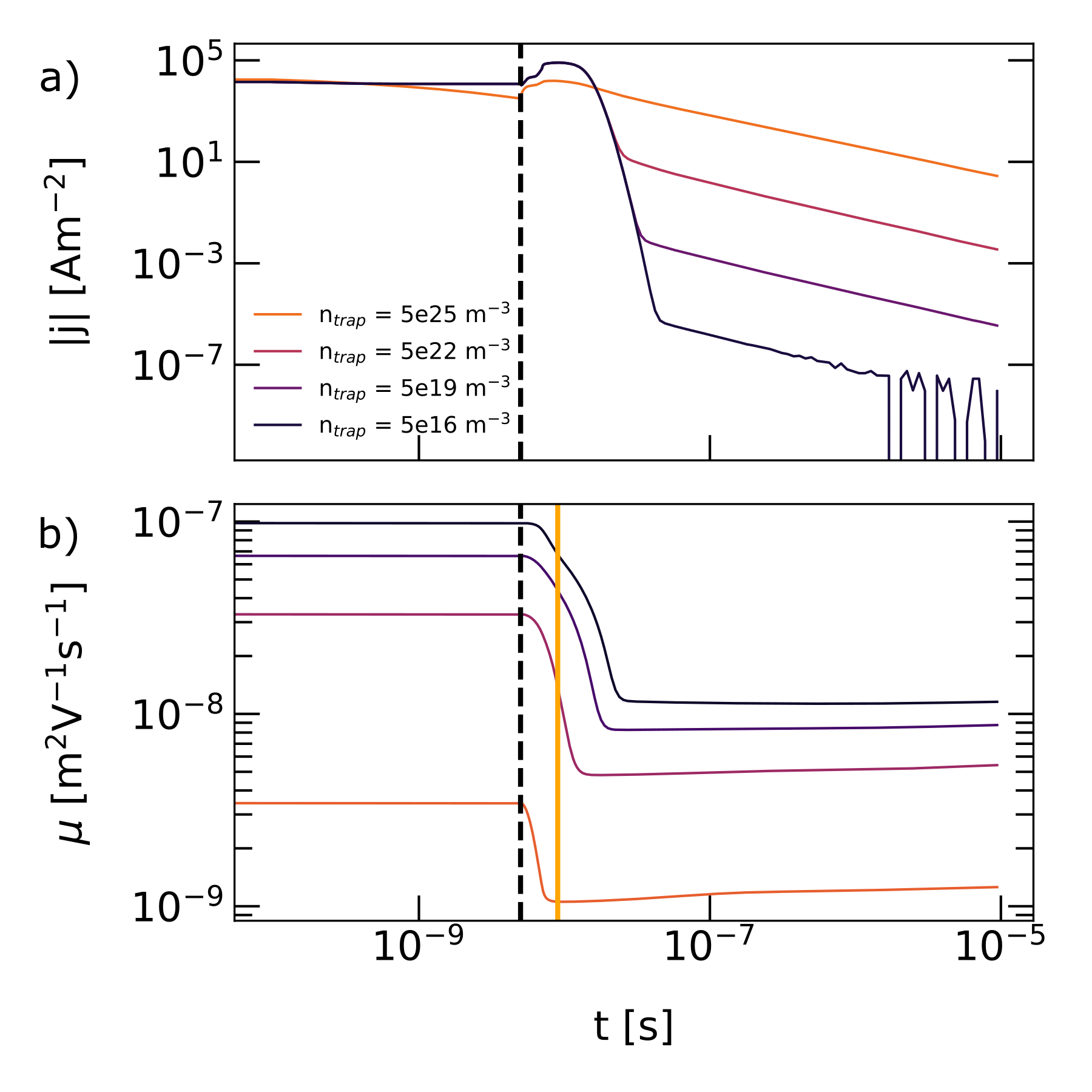}
    \caption{a) Comparison of current transients for different trap state densities and voltages. High trap state densities show a long lasting, high extraction current. b) Effective mobility during the TDCF experiment. The vertical lines are guides to the eye. The black dashed line indicates the time at which extraction starts and the orange line represents the time at which the applied voltage has finished ramping to $V_{col}$ }
    \label{fig:mobility_current_transients_paper}
\end{figure}

Figure \ref{fig:mobility_current_transients_paper}a depicts four TDCF transients with different densities of traps.  A trap density of $5\times10^{25}~\text{m}^{-3}$ is considered very disordered while a trap density of $5\times10^{16}~\text{m}^{-3}$ is considered quite trap free. It can be seen that for a device with many traps the TDCF transient slowly drops off due to the long time it takes for trapped carriers to be extracted due to their slow release from the trap states. In contrast, a TDCF transient from a device with few trapped states rapidly drops off as there are no trap states preventing the extraction of carriers. Thus, a simple test to determine if a device contains a significant number of trap states is to see if the TDCF transient decays rapidly or gradually.

The change in effective mobility calculated using Eqn.~(\ref{eq:paper_mueff}) is plotted in Fig.~\ref{fig:mobility_current_transients_paper}b. It can be seen that due to the changing ratio of free to trapped carriers, the average mobility within the device changes during the transient. An interesting question to ask is: Which mobility do we extract with the TDCF-method if the mobility is not constant? Do we measure the mobility at the start of the transient or the mid-point?

\begin{figure}
    \centering
    \includegraphics[width = \linewidth]{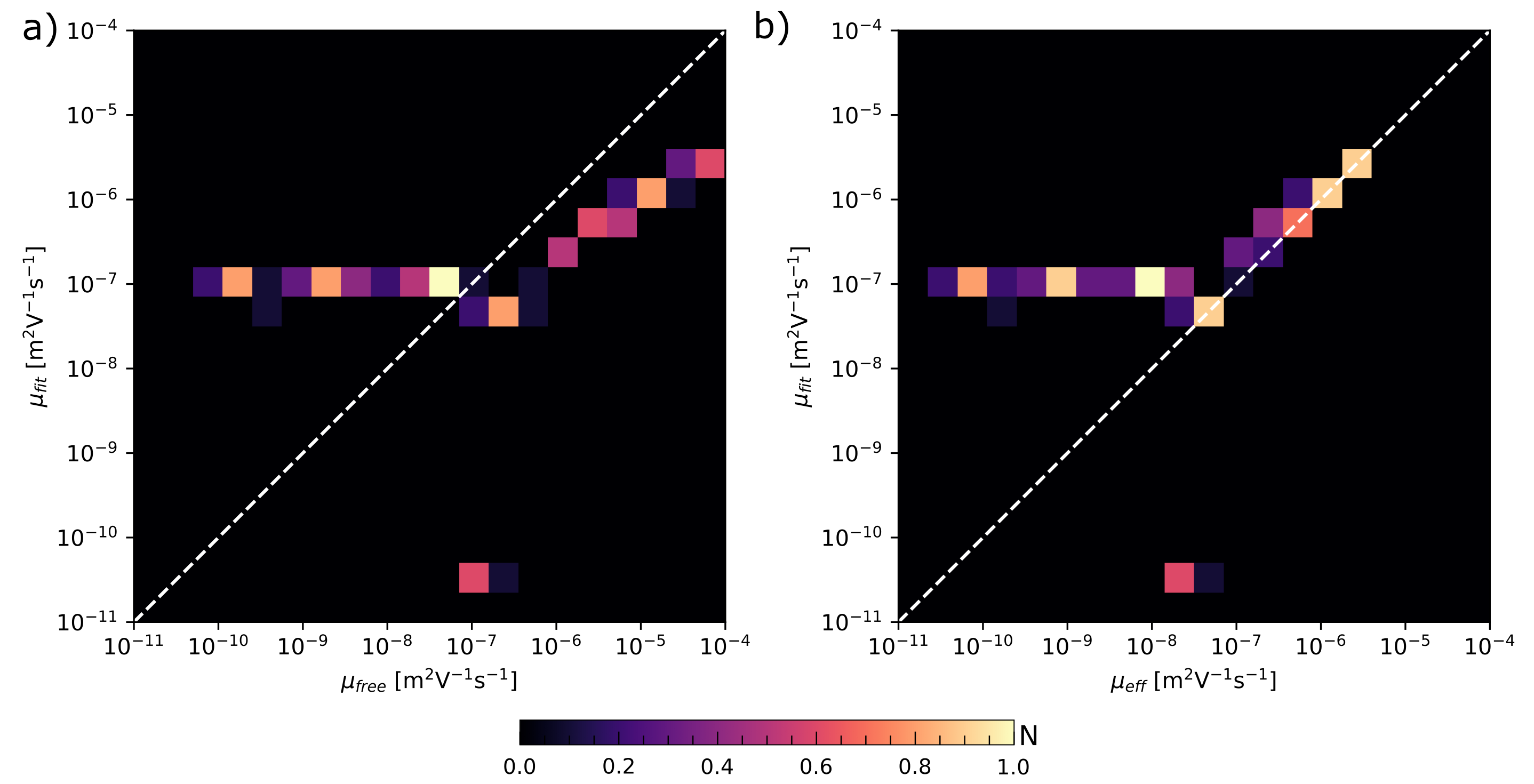}
    \caption{a) Comparison of free carrier mobility used to generate TDCF transients and the value of mobility extracted using the fitting method for a device with traps; b) Comparison of averaged carrier mobility just before the start of a TDCF transient plotted against the value of mobility extracted using the fitting method for a device with traps.}
    \label{fig:mobility_trap_vs_notrap_paper}
\end{figure}

Figure \ref{fig:mobility_trap_vs_notrap_paper}a plots the symmetric free carrier mobility used in the device model against that extracted from the transient using the TDCF fitting method. It can be seen that for low free carrier mobility values under $1 \times 10^{-7}~\text{m}^{2} \text{V}^{-1} \text{s}^{-1}$ TDCF struggles to extract accurate values of mobility, this is because multiple trapping and release prevents the carriers from leaving the device making extraction trap limited.  Once a device has a higher free carrier mobility, and a narrow distribution of extraction times, enough carriers can escape the device quickly enough for TDCF to start to work. However, we see an underestimation of the mobility by around half an order of magnitude. 
Figure~\ref{fig:mobility_trap_vs_notrap_paper}b is identical to Figure~\ref{fig:mobility_trap_vs_notrap_paper}a except that the effective mobility from Eqn.~(\ref{eq:paper_mueff}) is plotted on the x-axis.

Therefore, in summary, a high extraction current density at long times should be seen as a warning sign for a significant numbers of trap states. In such cases, care should be taken when interpreting mobility values -- especially if they are low -- and other techniques such as space-charge-limited current, frequency domain measurements or simulation should be used to verify the results.

\pagebreak
\clearpage

\section{Recombination rates}

In general, there are four common pathways for recombination in solar cells. They are free-to-free recombination which will have a recombination order of two, and free-to-trapped recombination which has a recombination order of one or larger.  In disordered systems, as they are many more trapped than mobile carriers, the latter mechanism will usually dominate the bulk recombination rate.
However, surface recombination at the contacts will push the recombination order towards one. Identifying the recombination process with the recombination order can give important information on how to minimise losses in a solar cell device \cite{aelm.201700505}.

\begin{figure}
    \centering
    \includegraphics[width = 0.7\linewidth]{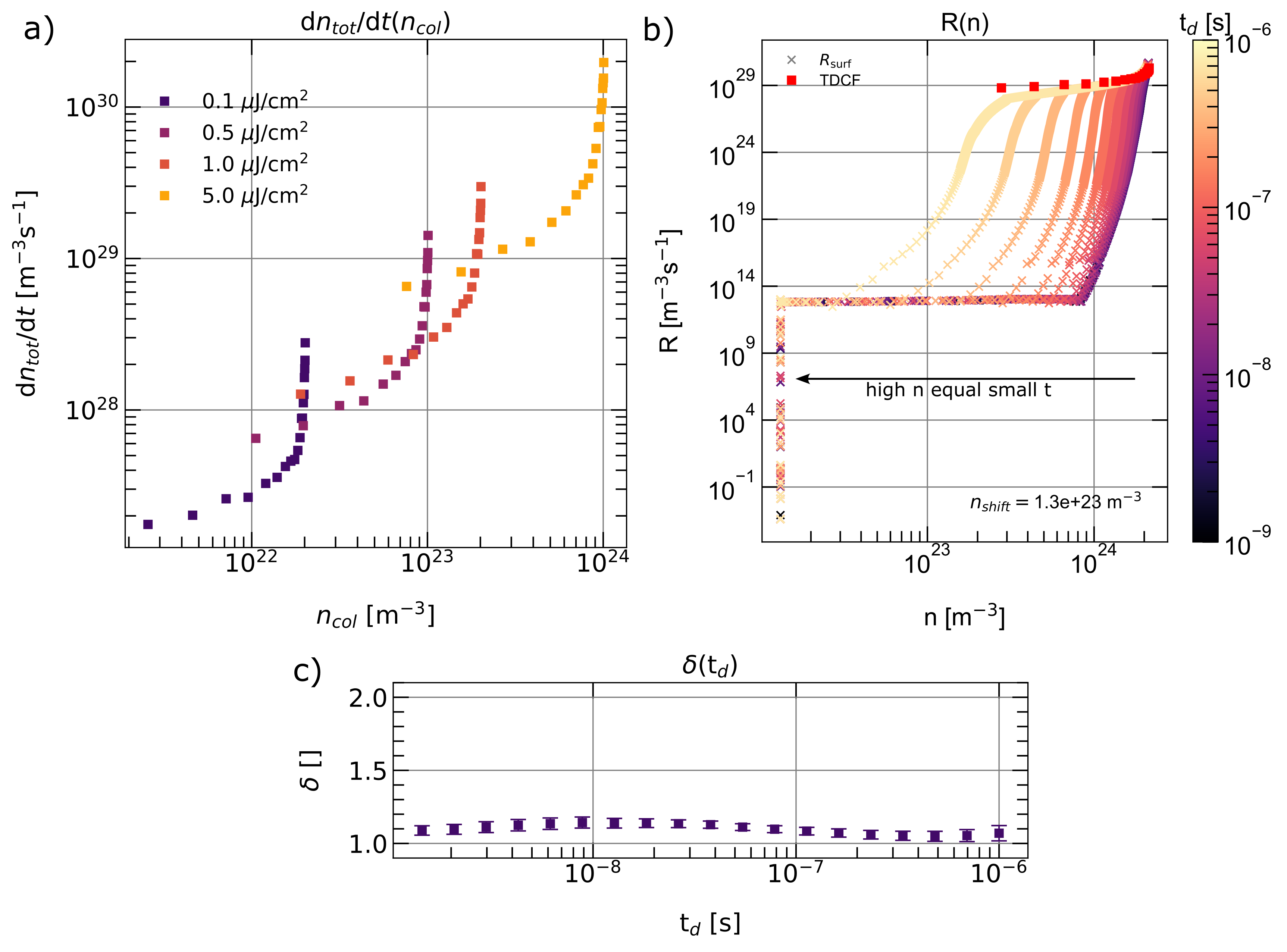}
    \caption{a) Recombination rates extracted from simulated TDCF transients for a device with no trap states; b) The recombination rate experienced by the device during charge carrier extraction for a range of delay times at a light intensity of $I = 0.1\unit{{\mu J}{cm^{-2}}}$ ; c) Time dependent recombination order $\delta(\td)$ for the device without traps, this is calculated by taking the gradient of the graph a for points collected at the same delay time.}
    \label{fig:recombination_no_traps}
\end{figure}

Recombination can be probed with TDCF by varying the delay time $t_{d}$. The integrated photo-current is equal to the charge extracted from the device. By plotting the differential change of the extracted charge carrier density $n_{tot}$ with respect to the delay time, a recombination rate can be extracted. If the recombination rate is plotted over the charge carrier density $n_{col}$ that survived recombination during the delay time it is possible to extract the recombination order. The recombination order $\delta$ is extracted from the slope of

\begin{equation}\label{eq:method_transittime}
    \delta=\frac{\mathrm{d}\ntot(\ncol)}{\mathrm{d}t}~,
\end{equation}
in a log-log-plot. For a given illumination intensity recombination rates at high charge carrier densities correspond to small delay times. Figure \ref{fig:recombination_no_traps}a shows the intensity dependent recombination rates for a device without trap states.

One can see in Figure \ref{fig:recombination_no_traps}a that each coloured data set consists of a two distinct regions, a gradually rising region followed by an inflection point and then a very steep region. In our device without trap states the very steep region probably implies surface recombination at the contacts \cite{Kniepert2019Chargeextraction, wuerfel_unmuessig2018apparentfielddependence}. 
To verify that surface recombination was indeed the dominating process, we picked one illumination intensity from Figure \ref{fig:recombination_no_traps}a ($I = 0.1\unit{{\mu J}{cm^{-2}}}$) and recorded the recombination rate $R$ and the charge density $n$ as a function of time during the TDCF transient. We repeated this for all the transients which make up the $I = 0.1\unit{{\mu J}{cm^{-2}}}$ data set. The result is plotted as multi-coloured crosses in Figure
\ref{fig:recombination_no_traps}b. The colours represent the delay time $t_d$. Thus darker colours represent transients extracted at later times. Plotted on the same figure are red squares, these represent recombination rates directly extracted from the simulated TDCF transients, as one would do if performing the experiment for real.

This enables a direct comparison between the measured recombination rate through TDCF and the time resolved recombination rates inside the device. It can be seen that the recombination rate within the device at the onset of the extraction exactly matches the surface recombination obtained from the TDCF transients.

Experimentally, surface recombination is detected by plotting the recombination orders at equal delay times for all the intensities. In practice this is done by fitting a straight line to the points in Figure \ref{fig:recombination_no_traps}a which have the same delay time. This will result in one point per light intensity being used for the fit each time.  The results from doing this can be seen in Figure \ref{fig:recombination_no_traps}c as the recombination is of the order $\delta(\td) = 1$ we can see that surface recombination is dominant in this device.

\begin{figure}[H]
    \centering
    \includegraphics[width = 0.7\linewidth]{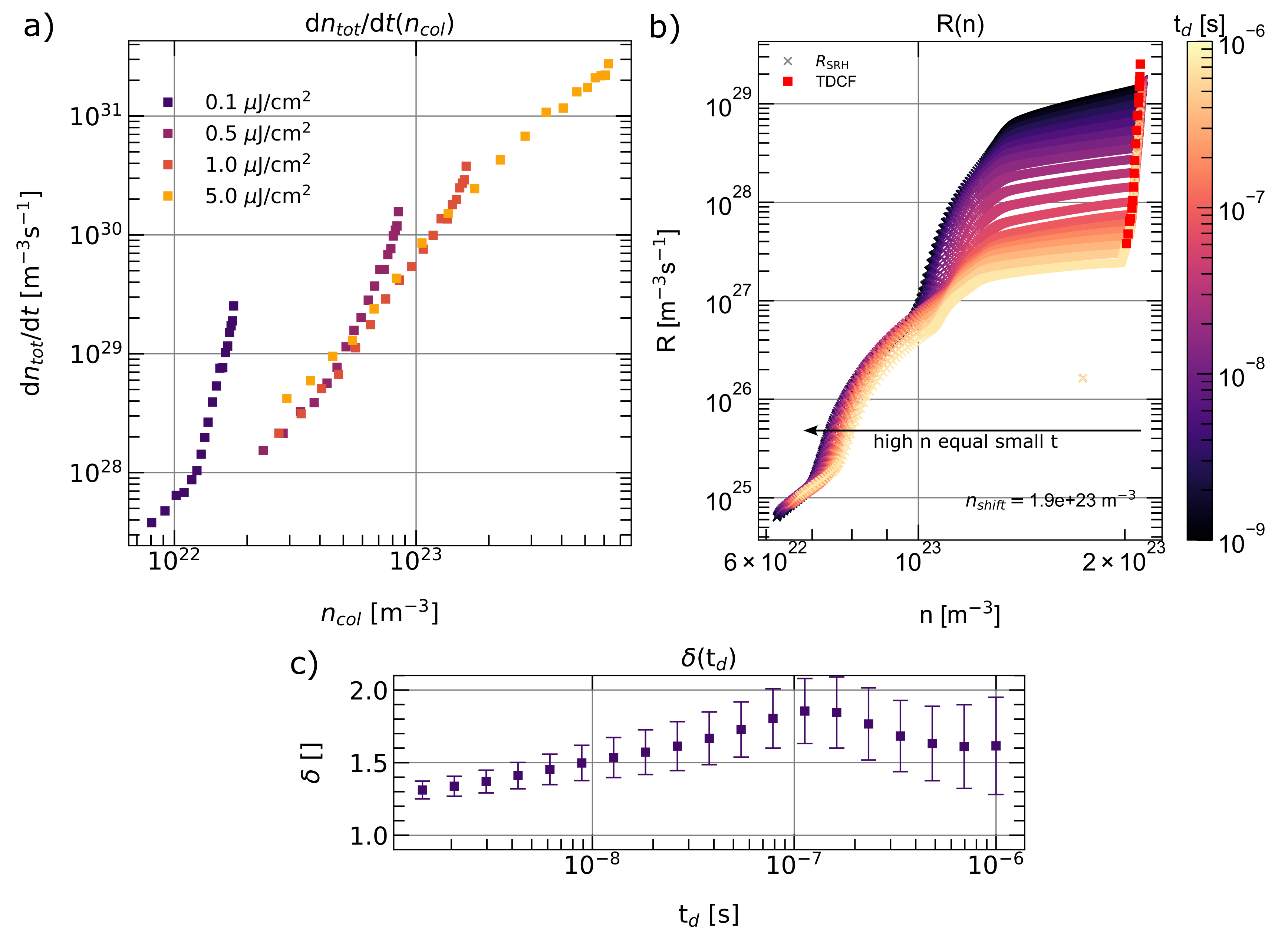}
    \caption{a) Recombination measured from virtual TDCF-experiments for a device with traps $N_\mathrm{trap} = 10^{25}\unit{m^{-3}}$; b)  Comparison of recombination rates extracted from TDCF experiments at $I = 0.1\unit{{\mu J}{cm^{-2}}}$ plotted against the internal $R_{SRH}$ recombination rate which includes carrier trapping as well as recombination. c) Extracted recombination order from a.}
    \label{fig:recombination_with_traps}
\end{figure}

We now turn our attention to a device with energetic disorder. Figure \ref{fig:recombination_with_traps}a plots the recombination rates extracted from virtual TDCF transients from a device with a trap density of $N_\mathrm{trap} = 10^{25}\unit{m^{-3}}$. For low light intensities the curves look like the case without traps in that there is initially a gradual rise in recombination rate followed by an inflection point.

Hofacker and Neher \cite{Hofacker_Neher2017dispersiverecomb} show that the steep parts of the curves which give high initial recombination orders in Fig.~\ref{fig:recombination_with_traps}a are due to fast initial thermalisation, which leads to rapidly dropping recombination rates while the charge carrier density changes only a little. Once the initial thermalisation is over, a quasi-equilibrium establishes and an asymptotic value of the recombination order establishes. As optical excitation intensities increase recombination rates become higher and dominate thermalisation.

Figure \ref{fig:recombination_with_traps}b plots the recombination rates extracted from the TDCF transients at low light intensity $I = 0.1\unit{{\mu J}{cm^{-2}}}$ as red squares. The crosses represent the $R_{SRH}$, which represents loss of free carriers though SRH recombination and trapping. $R_{SRH}$ was plotted within the TDCF transients for different delay times $t_{d}$.  High values of carrier density represent early times, while low values of carrier density represent late times.  It can be seen that the red squares sit to the very right hand side of the graph. At this light intensity we found $R_{SRH}$ dominated by trapping rather than recombination. Thus we concur with the results of Hofacker and Neher\cite{Hofacker_Neher2017dispersiverecomb}.

Figure \ref{fig:recombination_with_traps}c plots the extracted recombination order from \ref{fig:recombination_with_traps}a as a function of delay time.  A high initial recombination order with a time dependent recombination order $\delta \neq 1.0$ is an indicator for carrier thermalization and therefore an indication of trap states.

At early delay times, a recombination order of around 1.4 can be seen. Hofacker and Neher link early time carrier relaxation order to Urbach energy and temperature using the equation

\begin{equation}
    \delta_\mathrm{exp} = 1 + E_\mathrm{U}/k_\mathrm{B}T~.
\end{equation}

If we substitute in the value of Urbach energy ($0.1 \unit{eV}$) and temperature ($293\unit{K}$) used in our simulations this leads to a recombination order of $\delta_\mathrm{exp} = 1.4$ which can be observed in Figure \ref{fig:recombination_with_traps}c.

\section{Conclusion}

The charge extraction technique TDCF is capable of extracting charge carrier mobility by analysing the extraction current after applying a negative collection bias. The widely used method of fitting the initial photo-current decay (fit-method) is able to correctly determine the mobility of the faster charge carrier specimen in the case of negligible trap states and asymmetric charge carrier mobilities. To complement the fit-method a second approach to detect the slower charge carriers is proposed. By integrating the current transient to the time when 90\% of charge carriers are extracted $t_d$ can be calculated and the mobility of the slower carrier to be determined.
When trap states can not be neglected, TDCF is not always able to give reliable results for the mobility. The current transient at early times is dominated by charge carrier relaxation into trap states. During the course of the extraction, a high number of charge carriers reside in trapped states \cite{Hanfland2013CELIV}. These trapped carriers get freed step-by-step during the charge carrier extraction causing a long lasting high extraction current. The presence of latter is an indicator for the presence of high trap state densities. The extracted mobilities should be interpreted with caution.

By varying the delay time in TDCF it is possible to probe the recombination in a device and extract the recombination order depending on charge carrier density. Recombination orders greater than 1 for small delay times can occur and are often explained with the presence of surface recombination. It is shown that this is true for cases with no or negligible trap states but thermalisation of charge carriers in systems with high trap state densities can cause high initial recombination orders as well. Analysing the time dependent recombination order can give information on the origin of the high initial recombination orders and is therefore the second indicator of non-negligible trap states in TDCF experiments. We conclude that TDCF is a powerful method. However, it should be applied with caution or the support of simulations.

\bibliography{main}

\clearpage
\subsection{Supporting Material mobility}
\subsubsection{Asymmetric charge carrier mobility}

The mobilities used in Figures \ref{fig:mobility_asym_paper}a,b are; blue line: electron $1\times10^{-6} m^{2}V^{-1}s^{-1}$, hole $1\times10^{-6} m^{2}V^{-1}s^{-1} m^{2}V^{-1}s^{-1}$; orange line: Electron $3\times10^{-6} m^{2}V^{-1}s^{-1} m^{2}V^{-1}s^{-1}$ and $1\times10^{-6} m^{2}V^{-1}s^{-1}$; and green line $1\times10^{-5} m^{2}V^{-1}s^{-1}$ and $1\times10^{-6} m^{2}V^{-1}s^{-1}$

\Cref{fig:mobility_asym_paper} shows the extraction part of a simulated TDCF current-transient. Hole and electron mobility are highly asymmetric ($\mux{e} = 10^{-4}\unit{\frac{m^2}{Vs}}$ and $\mux{h} = 10^{-6}\unit{\frac{m^2}{Vs}}$). As a consequence the current-transient has two peaks. The first one for the electron extraction, the second one for the hole extraction. The orange line shows the initial photo-current decay extrapolation to determine the extraction time for the first peak. In theory the same method could be applied for the second peak. But the initial photo-current decay is less ambiguous for later times due to charge carrier diffusion processes. In a less extreme case the two peaks can overlap, making the fitting process even more unambiguous.

\begin{figure}[H]
    \centering
    \includegraphics[width = 0.5\textwidth]{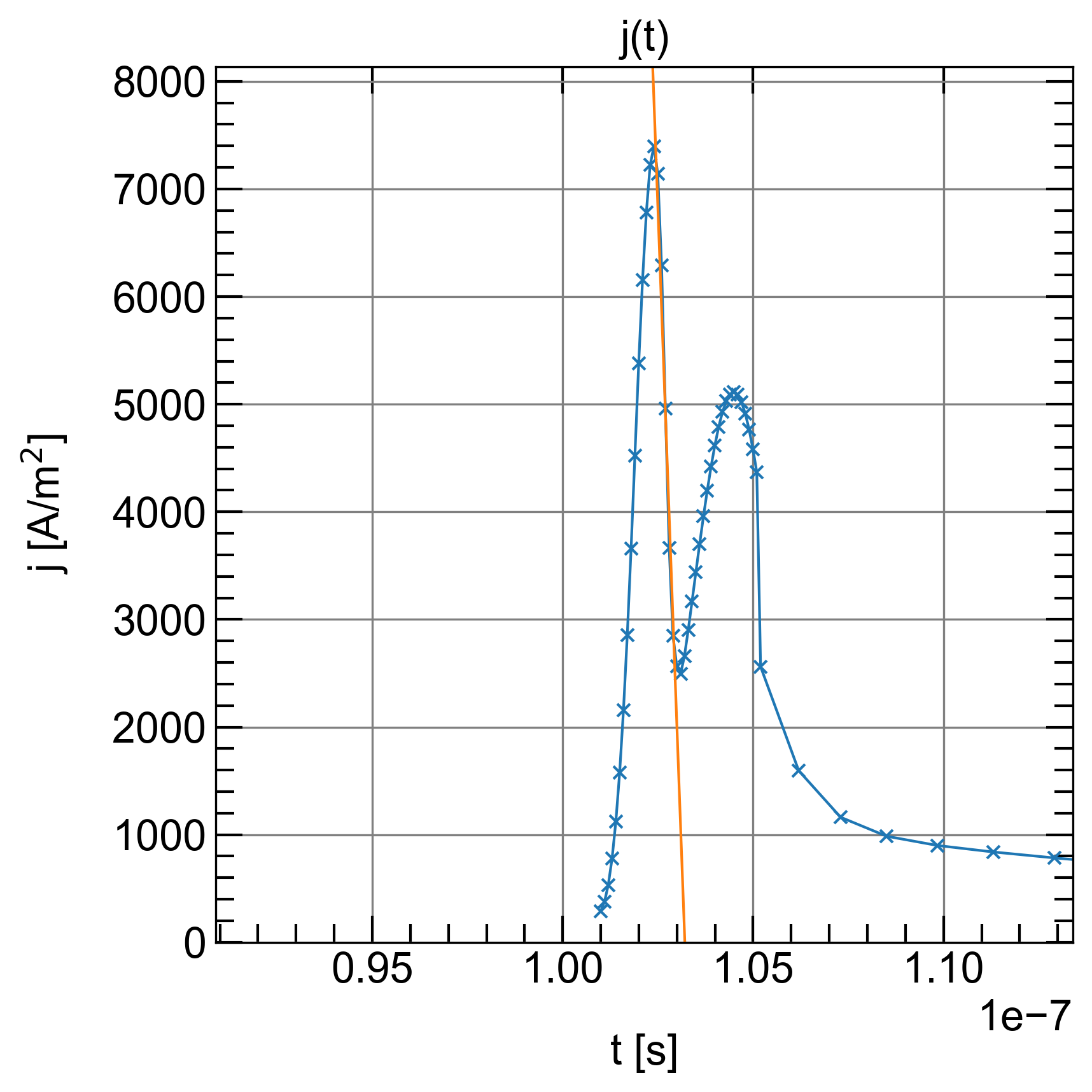}
    \caption{Transient of the photo current for highly asymmetric charge carrier mobilities $\mux{e} = 10^{-4}\unit{\frac{m^2}{Vs}}$ and $\mux{h} = 10^{-6}\unit{\frac{m^2}{Vs}}$.}
    \label{fig:mobility_asym_transient_paper}
\end{figure}

\subsubsection{RC-time limitation of mobility extraction}

In the main body of the paper, the confusion matrices in Fig. 2 c) and d) show slight derivations from the ideal case (diagonal dashed line) for high charge carrier mobilities. This is due to the RC-time constant of the solar cell \cite{Kniepert2017EffectOT}.
The contacts of the solar cell act as a capacitor that is charged when a bias in the TDCF experiment is applied to the device. The RC-time can be extracted from the simulated TDCF current-transients by fitting the current transient in the dark with the following two equations for the rise/fall of the current.

\begin{figure}[H]
    \begin{subfigure}[t]{0.49\linewidth}
        \centering
        \includegraphics[width = \textwidth]{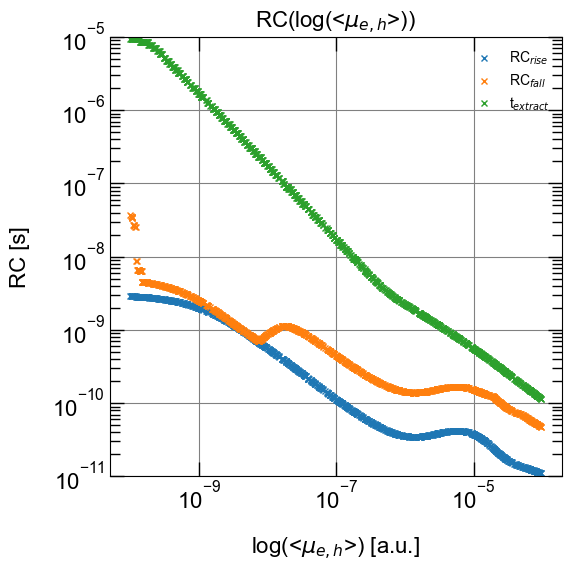}
        \caption{Comparison between determined extraction time and RC-time depending on mobility. It gets evident, that RC-time and extraction time have the same order of magnitude for high input mobilities. Thus RC-time correction of the extraction time gets necessary.}
        \label{fig:mobility_free_rc_time_paper}
    \end{subfigure}
    \hspace{\xvert\textwidth}
    \begin{subfigure}[t]{0.49\linewidth}
        \centering
         \includegraphics[width = \textwidth]{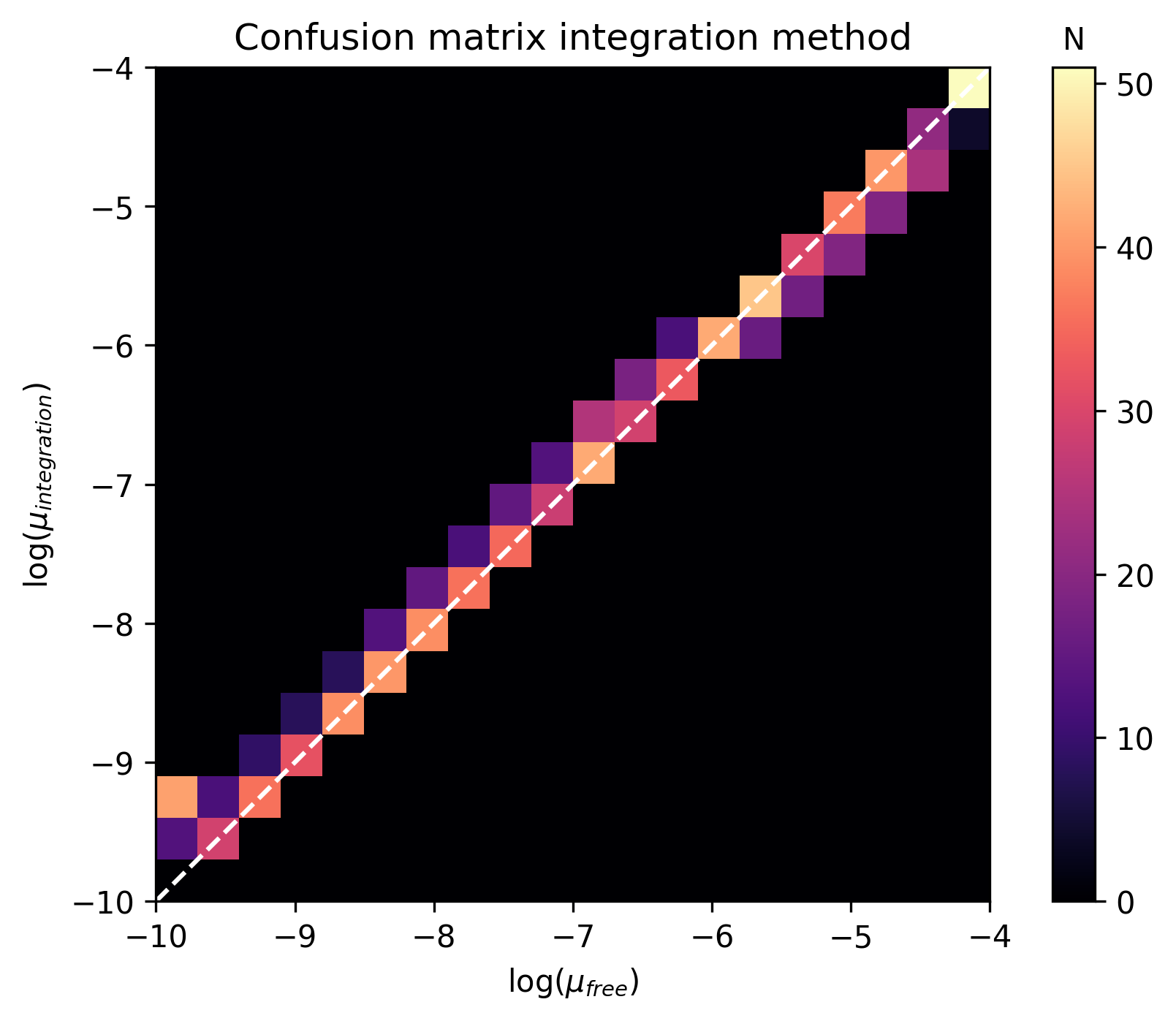}
         \caption{Extracted mobility with correction for RC-time. The Deviations for high mobilities have vanished.}
        \label{fig:mobility_free_rc_corrected_paper}
    \end{subfigure}
    \caption{Influence of the RC-time on extracted mobility on the example of the integration method.}
    \label{fig:mobility_RC_paper}
\end{figure}

In \Cref{fig:mobility_RC_paper} a) show the calculated RC-times depending on charge carrier mobility and the extraction time calculated from the initial photo-current decay. It can be seen that the RC-times and the extraction times for high mobilities are in the dame order of magnitude. The charging of the capacitor shifts the current-peak to later times. If $RC \approx t_\mathrm{extract}$, the initial photo-current decay method highly overestimates the extraction time (therefore underestimates mobility). After correcting the extraction time for the RC-time the charge carrier mobility can be accurately determined, as is shown in \Cref{fig:mobility_RC_paper} b).

\subsubsection{Charge carrier trapping in TDCF current-transients}

\begin{figure}%[H]
    \centering
    \begin{subfigure}[t]{0.4\linewidth}
        \centering
        \includegraphics[width =\textwidth]{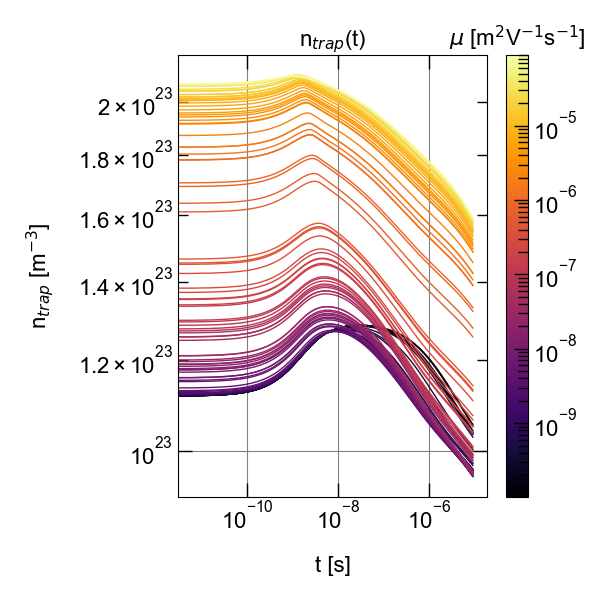}
        \caption{$n_{\mathrm{trap}}(t) $ for $n_{\mathrm{trap}}=5\cdot 10^{25}\unit{m^{-3}}$, note the increase in $n_{\mathrm{trap}}$ even after the delay time at $\td = 10^{-9}\unit{s}$ (this could be due to in-homogeneous charge carrier excitation)}
        \label{fig:mobility_trap_ntrap_paper}
    \end{subfigure}
    \hspace{\xvert\textwidth}
    \begin{subfigure}[t]{0.4\linewidth}
        \centering
        \includegraphics[width = \textwidth]{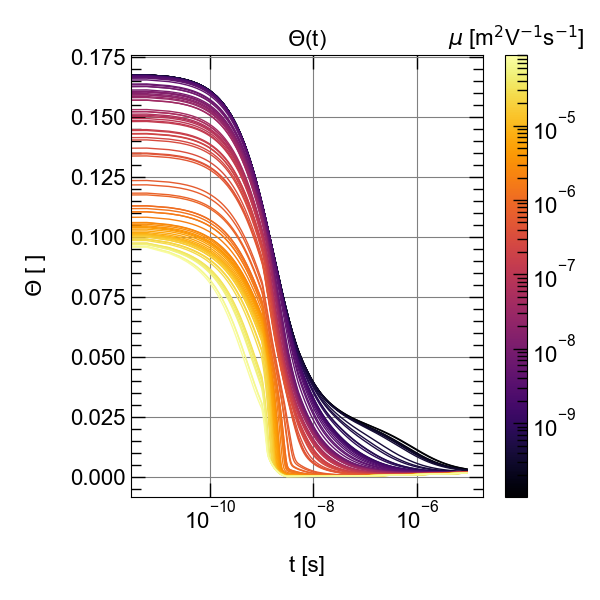}
        \caption{$\Theta(t) = \frac{n_{\mathrm{free}}}{n_{\mathrm{free}}+ n_{\mathrm{trap}}}$}
        \label{fig:mobility_trap_theta_paper}
    \end{subfigure}
    \begin{subfigure}[t]{0.4\linewidth}
        \centering
        \includegraphics[width = \textwidth]{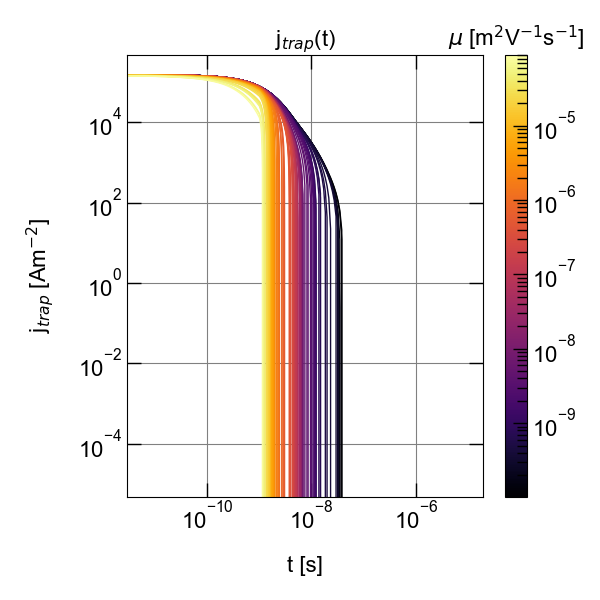}
        \caption{Current inside the device caused by trapping of charge carriers: $j_{\mathrm{trap}}(t) = T_{\mathrm{SRH}}\cdot \mathrm{e} \cdot L$}
        \label{fig:mobility_trap_jtrap_paper}
    \end{subfigure}
    \hspace{\xvert\textwidth}
    \begin{subfigure}[t]{0.4\linewidth}
        \centering
        \includegraphics[width = \textwidth]{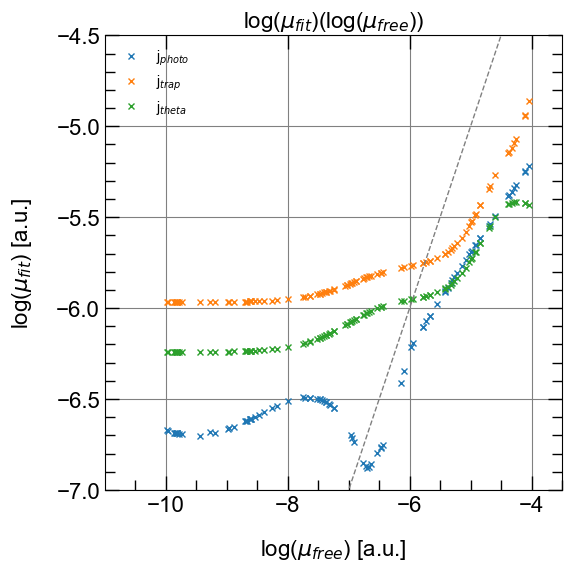}
        \caption{Comparison of determined mobility by linear photo-current decay and linear extrapolation of the time, all carriers are trapped (from $\Theta$(t) and $j_{\mathrm{trap}}$(t)). The behaviour of the mobility seems to be dominated by trapping effects for the time interval the photo-current decays linearly.}
        \label{fig:mobility_trap_mu_trap_paper}
    \end{subfigure}
    \caption{Dynamics of trapped charge carriers and their influence on the determined mobility.}
    \label{fig:mobility_trap_paper}
\end{figure}

In the following it will be reasoned why the process that causes the intial photo-current decay extrapolation to fail in Fig. 4 is charge carrier trapping.
In \Cref{fig:mobility_trap_paper} a) the simulated trap state density for different input mobilities is shown with respect to time during the TDCF experiment. Note the shift of the peak in trap state density.
This is [ossibly because slower charge carriers get moved through the device more slowly and as a consequence are more likely to get trapped in vacant trap states. \Cref{fig:mobility_trap_paper} b) shows the temporal evolution of $\Theta$ which is the fraction of free charge carriers to total charge carriers. It is apparent that charge carrier extractio is more efficient for high charge carrier mobilities. $\Theta(t)$ decays less slowly, because the slower charge carriers get trapped and de-trapped more frequently than the fast ones.
To visualise the influence of charge carrier trapping on the current transient, a trapping current is shown in \Cref{fig:mobility_trap_paper} c). It is calculated from the simulated SRH trapping rates $T_\mathrm{SRH}$:
\begin{equation}
    j_{\mathrm{trap}}(t) = T_{\mathrm{SRH}}\cdot \mathrm{e} \cdot L
\end{equation}
with the elementary charge $e$ and the device thickness $L$.
Both the trapping current and the $\Theta(t)$ are extrapolated to zero to determine a quasi extraction time. The results are compared with the extraction times from the original photo-current decay in \Cref{fig:mobility_trap_paper} d). Both approaches exhibit the same trend as the photo-current decay results. Therefore we conclude that charge carrier trapping is indeed the process that dominates the photo-current at high trap state densities in combination with low charge carrier mobilities.

\begin{figure}%[H]
    \centering
    \begin{subfigure}[t]{0.4\linewidth}
        \centering
        \includegraphics[width =\textwidth]{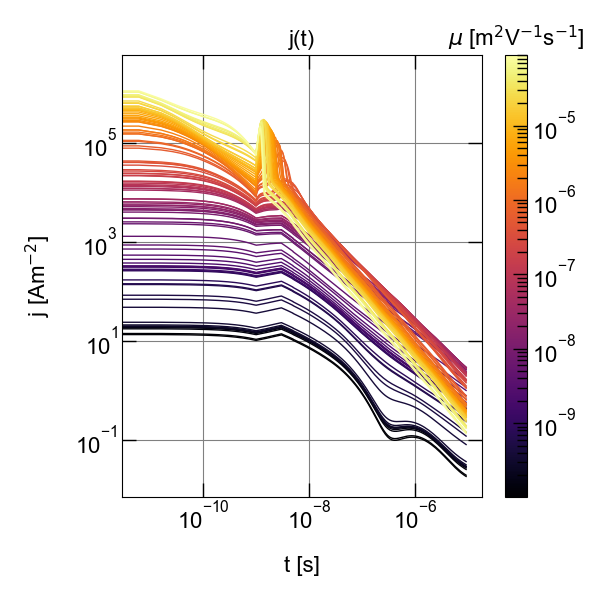}
        \caption{$j_{\mathrm{photo}}(t) $ for $n_{\mathrm{trap}}=5\cdot 10^{25}\unit{m^{-3}}$}
        \label{fig:mobility_notrap_jphoto_paper}
    \end{subfigure}
    \hspace{\xvert\textwidth}
    \begin{subfigure}[t]{0.4\linewidth}
        \centering
        \includegraphics[width = \textwidth]{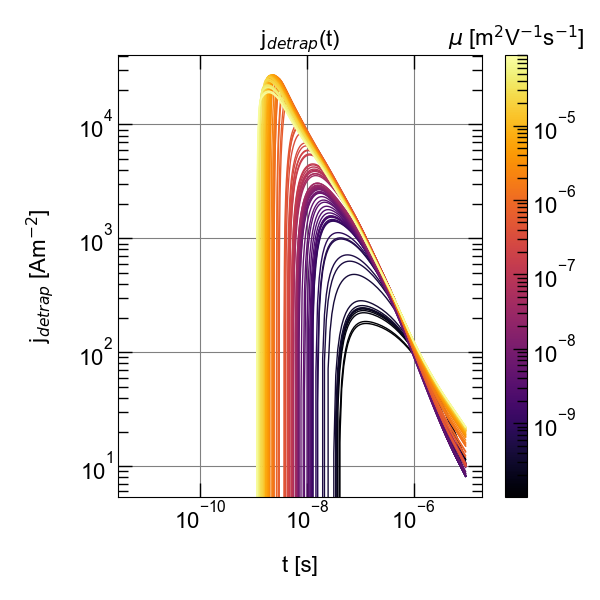}
        \caption{$j_{\mathrm{detrap}}(t) = -T_{\mathrm{SRH}}\cdot \mathrm{e} \cdot L$, calculated detrapping current density in the device}
        \label{fig:mobility_notrap_jdetrap_paper}
    \end{subfigure}
    \begin{subfigure}[t]{0.4\linewidth}
        \centering
        \includegraphics[width = \textwidth]{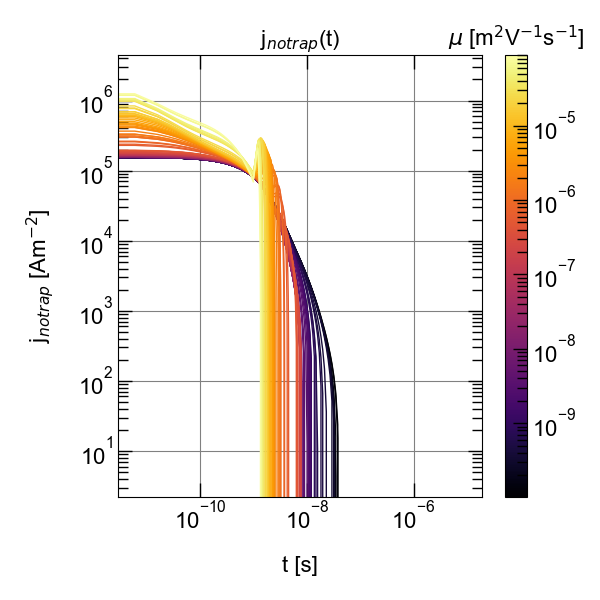}
        \caption{$j_{\mathrm{notrap}}(t) = j_{\mathrm{photo}}(t)- j_{\mathrm{trap}}(t)$, to exclude the influence of trapping, only the initial decay of the positive current is fitted.}
        \label{fig:mobility_notrap_jnotrap_paper}
    \end{subfigure}
    \hspace{\xvert\textwidth}
    \begin{subfigure}[t]{0.4\linewidth}
        \centering
        \includegraphics[width = \textwidth]{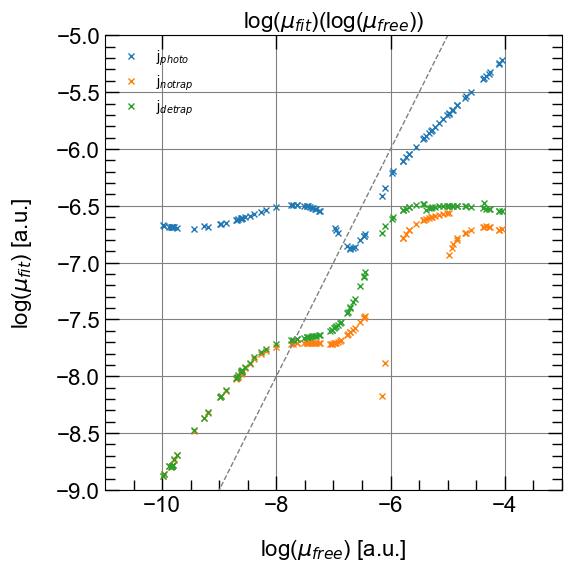}
        \caption{Comparison of determined mobility by linear photo-current decay and linear extrapolation of detrapping-current and photo-current without trapping effects. As one can see, the plateau for low mobilities seems to vanish, if the influence of trapping is excluded.}
        \label{fig:mobility_notrap_mu_notrap_paper}
    \end{subfigure}
    \caption{Dynamics of trapped charge carriers and their influence on the determined mobility.}
    \label{fig:mobility_notrap_paper}
\end{figure}

In the next step we try to artificially exclude the influence of trapping by subtracting the trapping-current from the photo-current in \Cref{fig:mobility_notrap_paper} a) to get the corrected current in \Cref{fig:mobility_notrap_paper} c):
\begin{equation}
    j_{\mathrm{notrap}}(t) = j_{\mathrm{photo}}(t)- j_{\mathrm{trap}}(t)
\end{equation}
The the current is again extrapolated towards $j=0$. The comparison with the results from the photo current in \Cref{fig:mobility_notrap_paper} d) show that the corrected current gives results much closer to the actual input mobilities.

\subsubsection{Energetic landscape influenced by trap states}

\begin{figure}%[H]
    \centering
    \includegraphics[width = 0.7\linewidth]{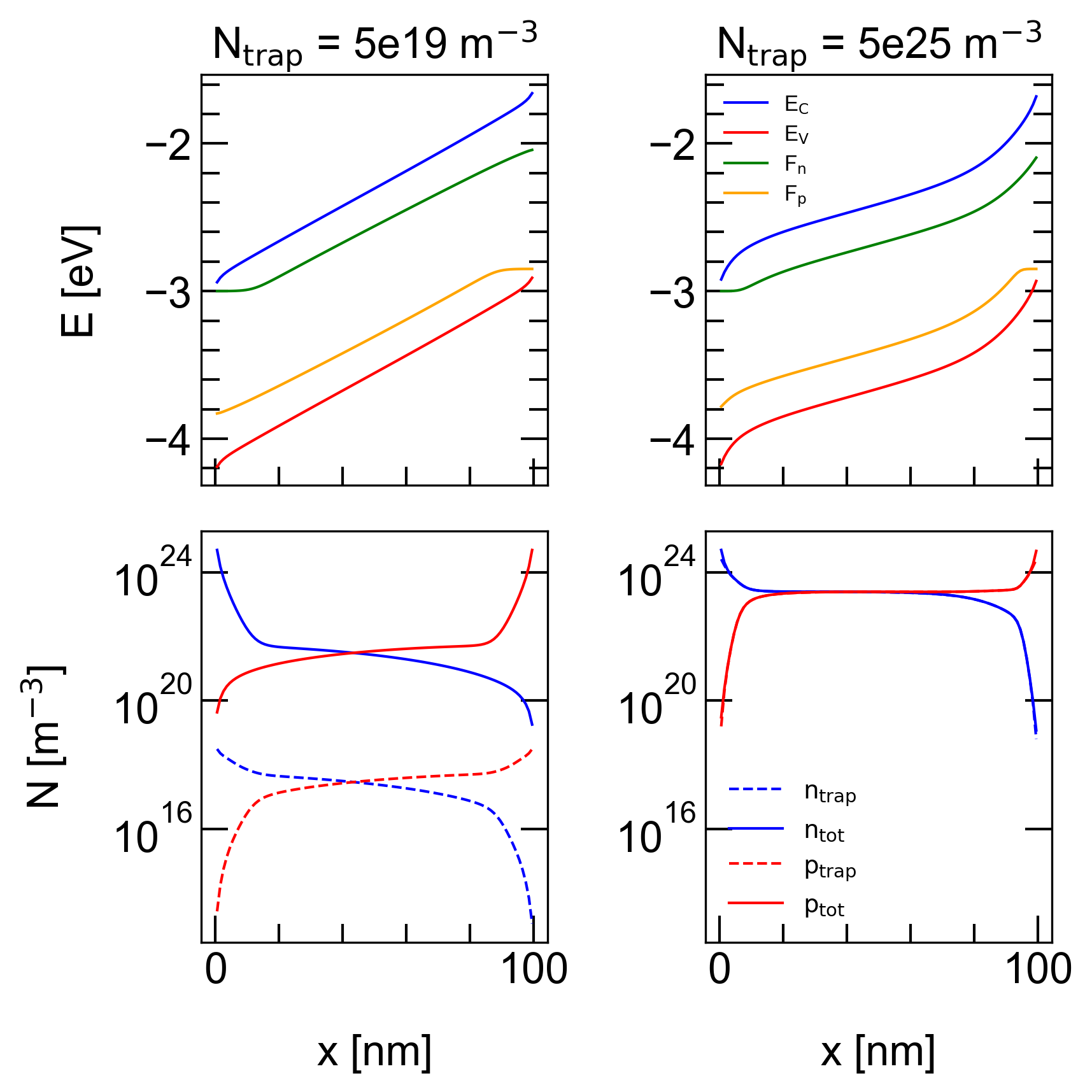}
    \caption{Free charge carrier density and trapped charge carrier density with the according band diagrams. For low trap state densities, free charges get extracted easily, after that only charge carriers from deep trap states are available. For high trap state densities,  $N_{\mathrm{free}} \approx N_{\mathrm{trap}}$ and charge carriers from shallow traps can get extracted easily.}
    \label{fig:mobility_band_paper}
\end{figure}

\Cref{fig:mobility_band_paper} depicts the energetic band structure and the charge carrier density spatially resolved for a device with low and with high trap state density. One can see that the conduction and valence band get highly bend by the trapped charge carriers. As many carriers are trapped and as a consequence are not mobile anymore, a space charge builds in the device causing a band bend. Organic materials often have high trap state densities due to their energetic disorder. Therefore idealised models like straight energy bands are nor valid. Organic semiconductor devices should always be simulated with trap states.

\clearpage
\subsection{Supporting material Recombination}

\subsubsection{Comparing simulated TDCF experiments with drift diffusion results}

\begin{figure}[H]
    \centering
    \begin{subfigure}[t]{0.4\linewidth}
        \includegraphics[width = \textwidth]{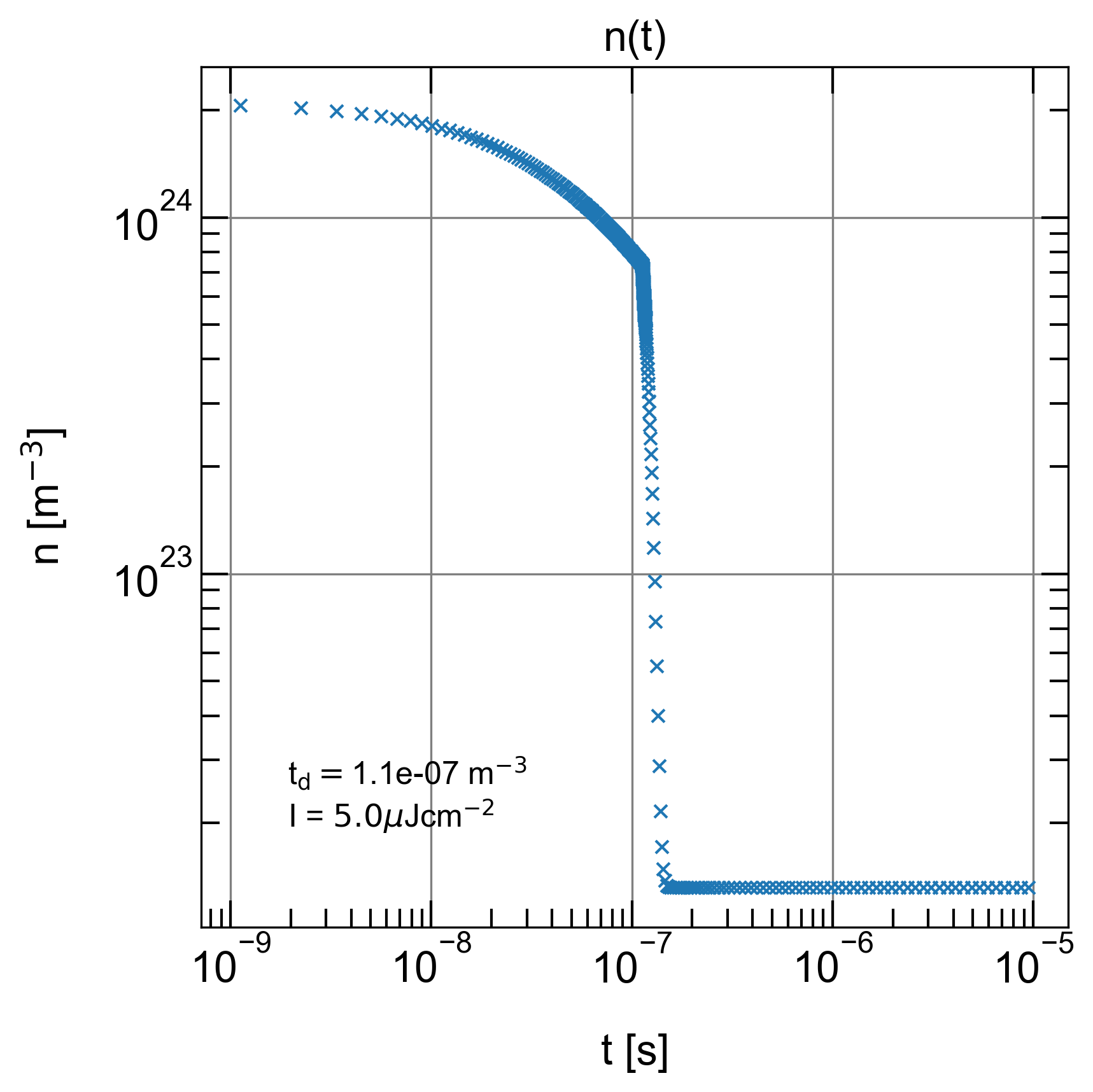}
        \caption{Total charge carrier concentration inside the device depending on time from drift-diffusion simulations.}
        \label{fig:tdcf_n(t)_paper}
    \end{subfigure}
    \hspace{\xvert\textwidth}
    \begin{subfigure}[t]{0.4\linewidth}
        \includegraphics[width = \textwidth]{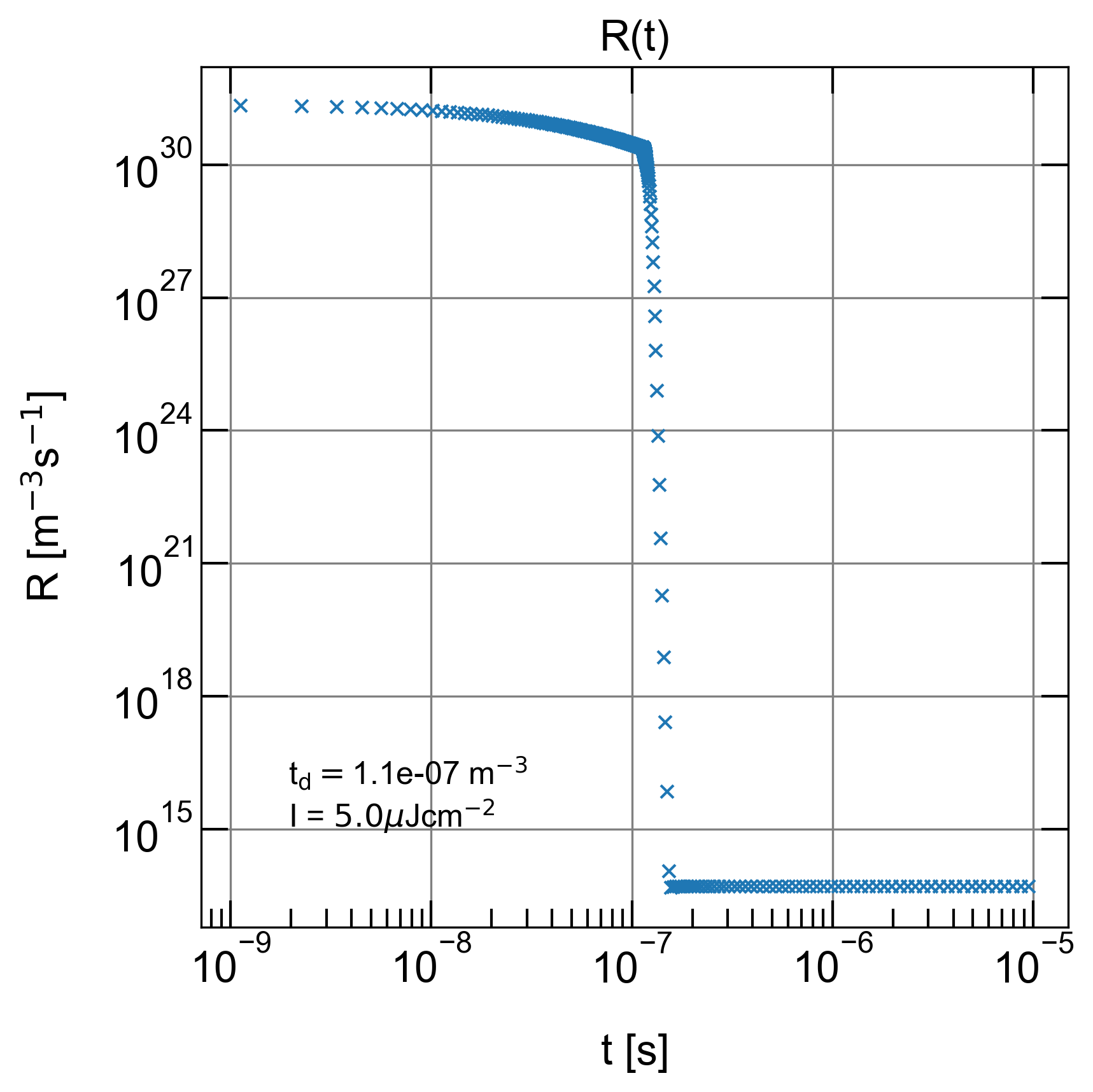}
        \caption{Recombination rates depending on time from drift-diffusion simulations.}
        \label{fig:tdcf_r(t)_paper}
    \end{subfigure}
    \begin{subfigure}[t]{0.4\linewidth}
        \includegraphics[width = \textwidth]{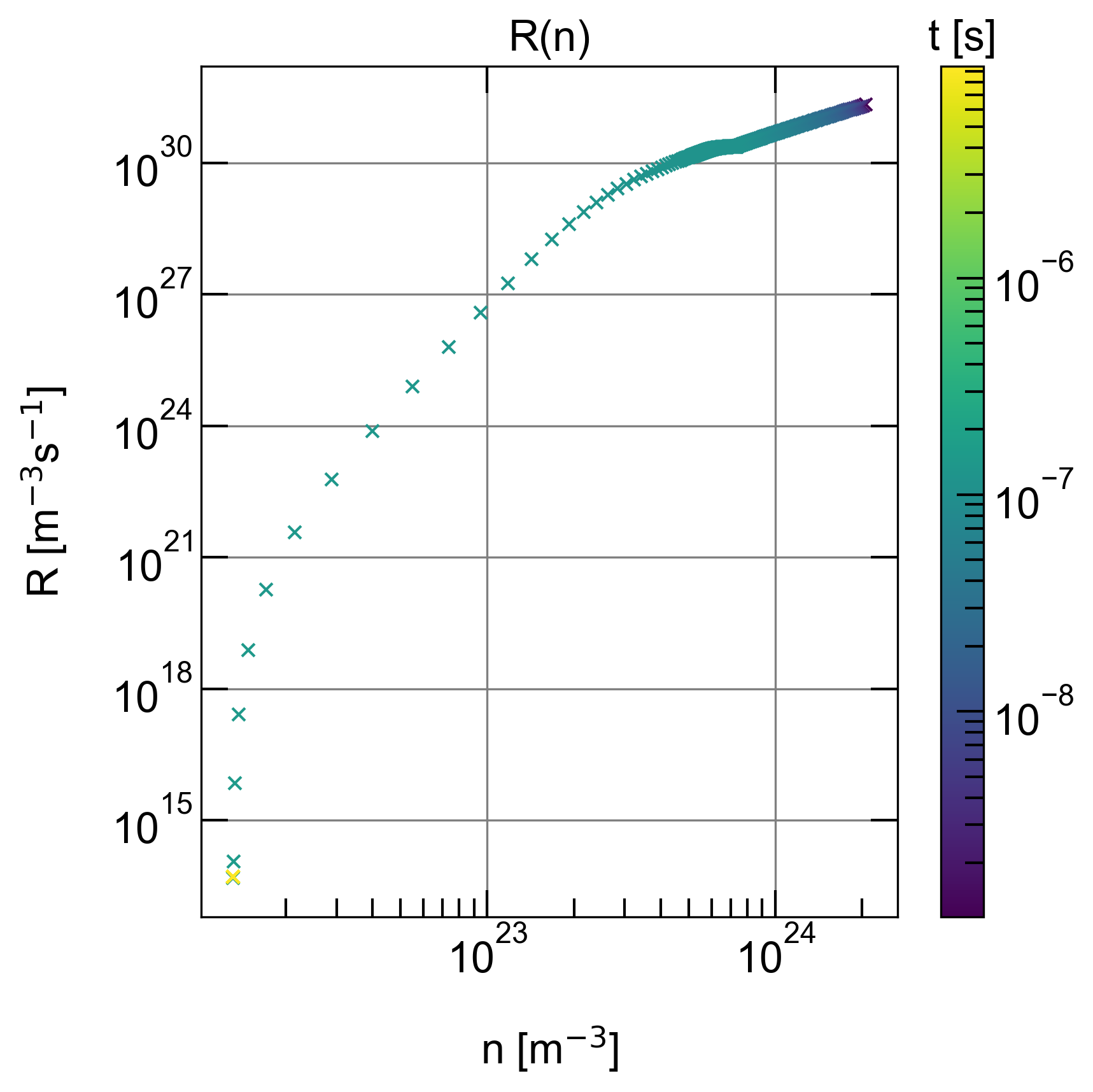}
        \caption{Recombination rate depending on charge carrier density by combining \cref{fig:tdcf_n(t)_paper} and \cref{fig:tdcf_r(t)_paper} for the delay time $\td = 1.1\cdot10^{-7}\unit{s}$ and laser intensity $I = 5.0\unit{\frac{\mu J}{cm^2}}$.}
        \label{fig:tdcf_r(n)_paper}
    \end{subfigure}
    \caption{The figures show how the recombination rates depending on charge carrier density are inferred from time dependent recombination rates and charge carrier densities determined by drift-diffusion simulations.}
    \label{fig:tdcf_transient_paper}
\end{figure}

To get a better understanding about the information gained from TDCF experiments, we compare the results from simulated TDCF-experiments (applying the usual data analysis to simulated current-transients) directly with the results from the drift diffusion model (parameters obtained during the simulation). The process is shown in \Cref{fig:tdcf_transient_paper}. The current density and recombination rate over the course of the simulated experiment can be seen in a) and b) respectively. By combining $R(t)$ and $n(t)$ as $R(n)$ we can compare the simulation parameters with the parameters extracted from TDCF. $R(n)$ is shown for one delay time in \Cref{fig:tdcf_transient_paper} c). The colour code visualises the time axis. In later comparisons process is done for all delay times in the TDCF experiment.

\subsubsection{Recombination rates for a device without trap states}

\begin{figure}[H]
    \centering
    \begin{subfigure}[t]{0.49\linewidth}
        \includegraphics[width = \textwidth]{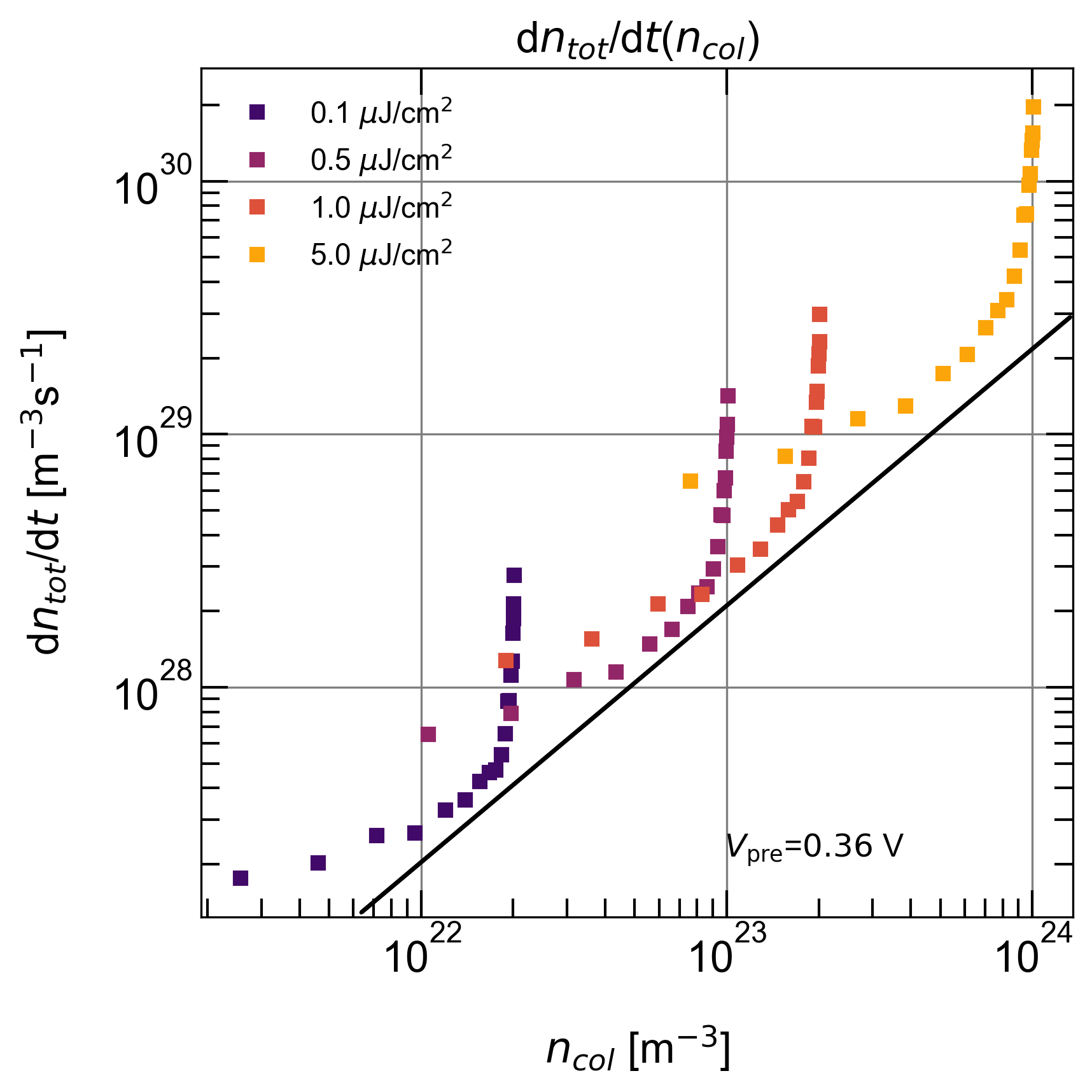}
        \caption{TDCF-simulation with parameters from \cref{tab:tdcf_surf_paper}. \textbf{Case 1} with low free-to-free recombination and no trap states. Solid lines are guides to the eye. Black: slope $\delta = 1$.}
        \label{fig:tdcf_tdcf_extract_paper}
    \end{subfigure}
    \hspace{\xvert\textwidth}
    \begin{subfigure}[t]{0.49\linewidth}
        \includegraphics[width = \textwidth]{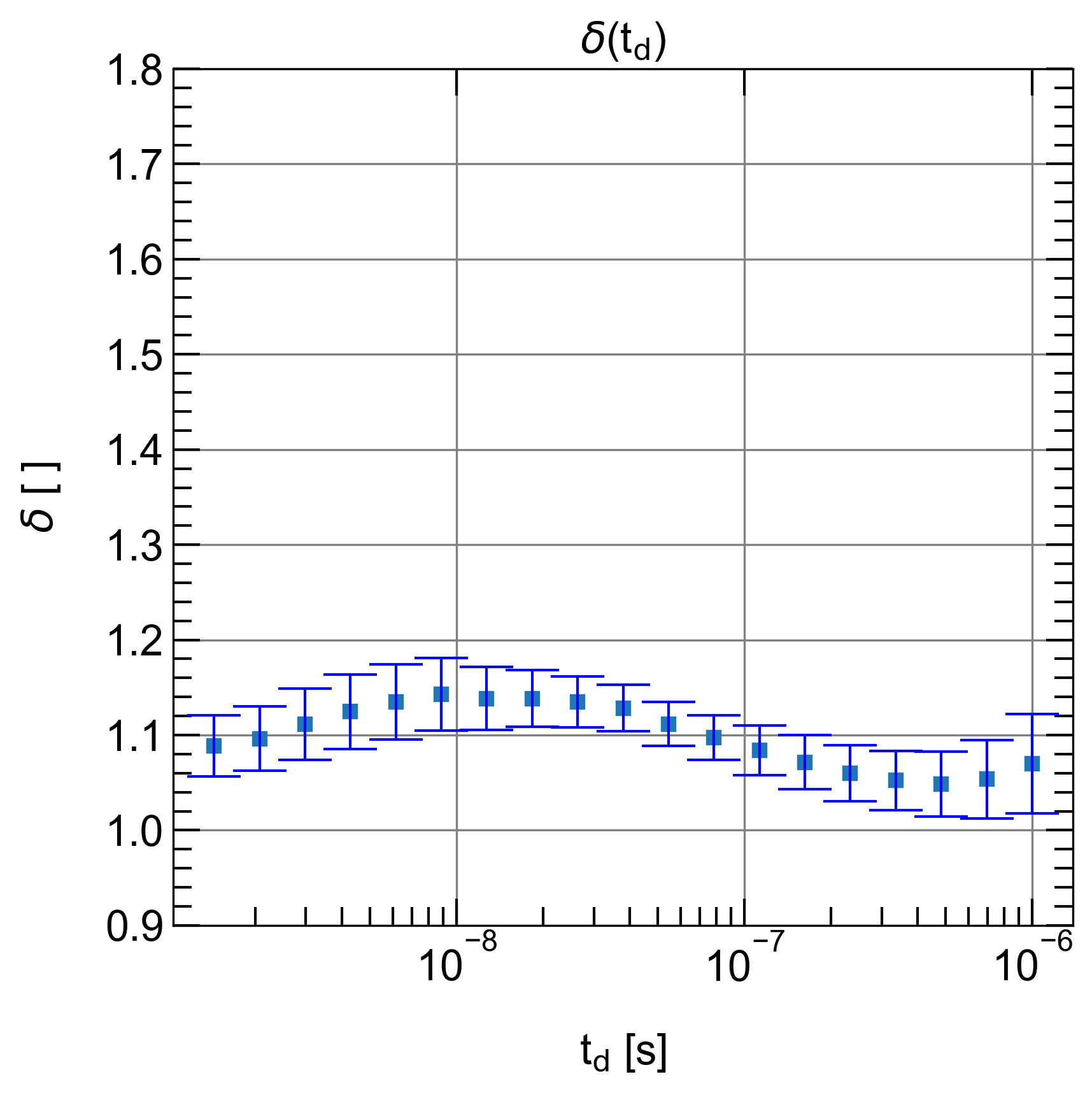}
        \caption{Recombination order $\delta (\td)$ inferred from linear fits of TDCF-data with the same delay time in a log-log-plot.}
        \label{fig:tdcf_lambda_extract_paper}
    \end{subfigure}
    \hspace{\xvert\textwidth}
    \begin{subfigure}[t]{0.49\linewidth}
        \includegraphics[width = \textwidth]{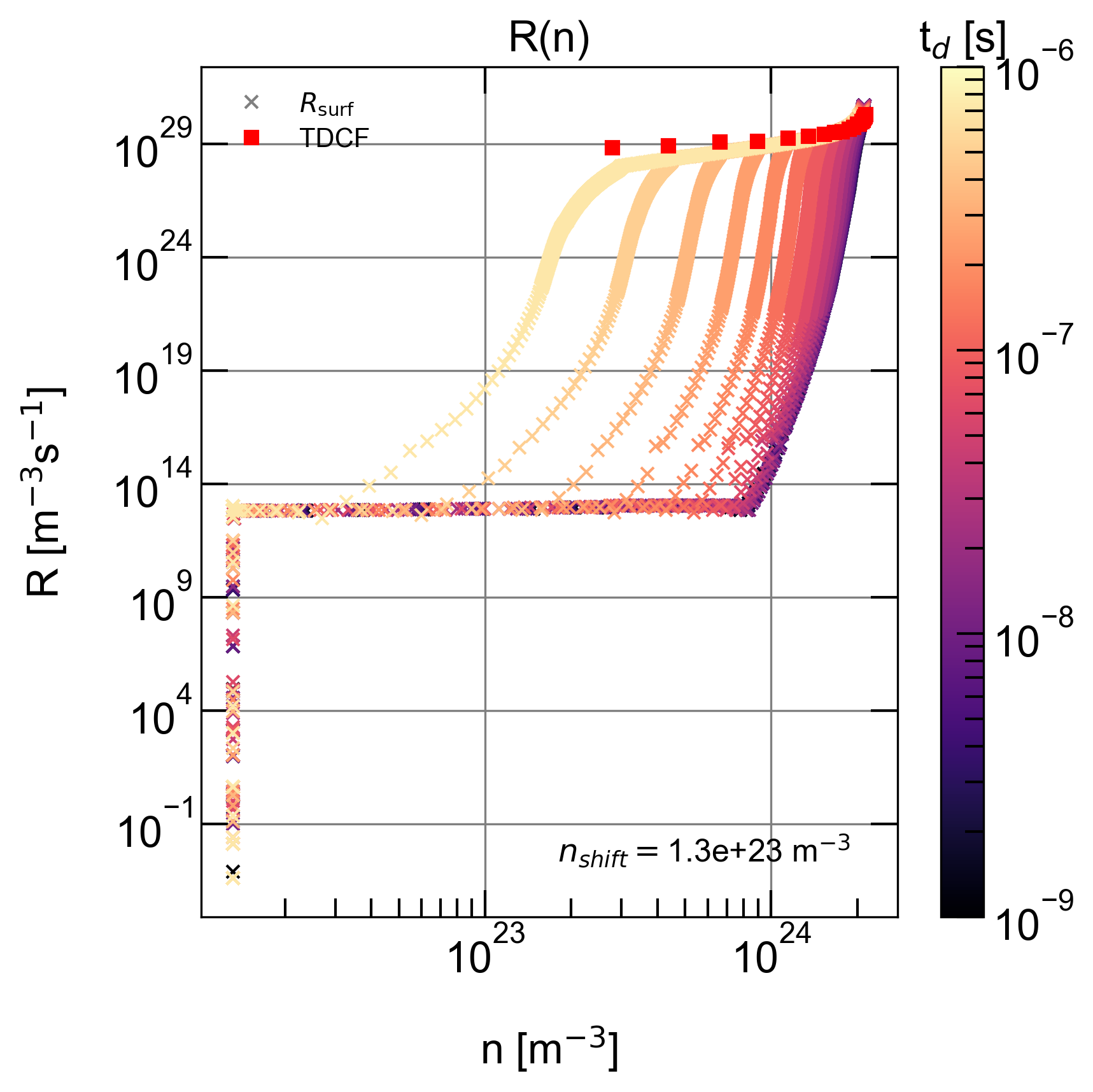}
        \caption{Comparison of TDCF-data with R(n): R is surface-recombination $R_{\mathrm{surf}}$.}
        \label{fig:tdcf_r(n)_extract_paper}
    \end{subfigure}
    \begin{subfigure}[t]{0.49\linewidth}
        \includegraphics[width = \textwidth]{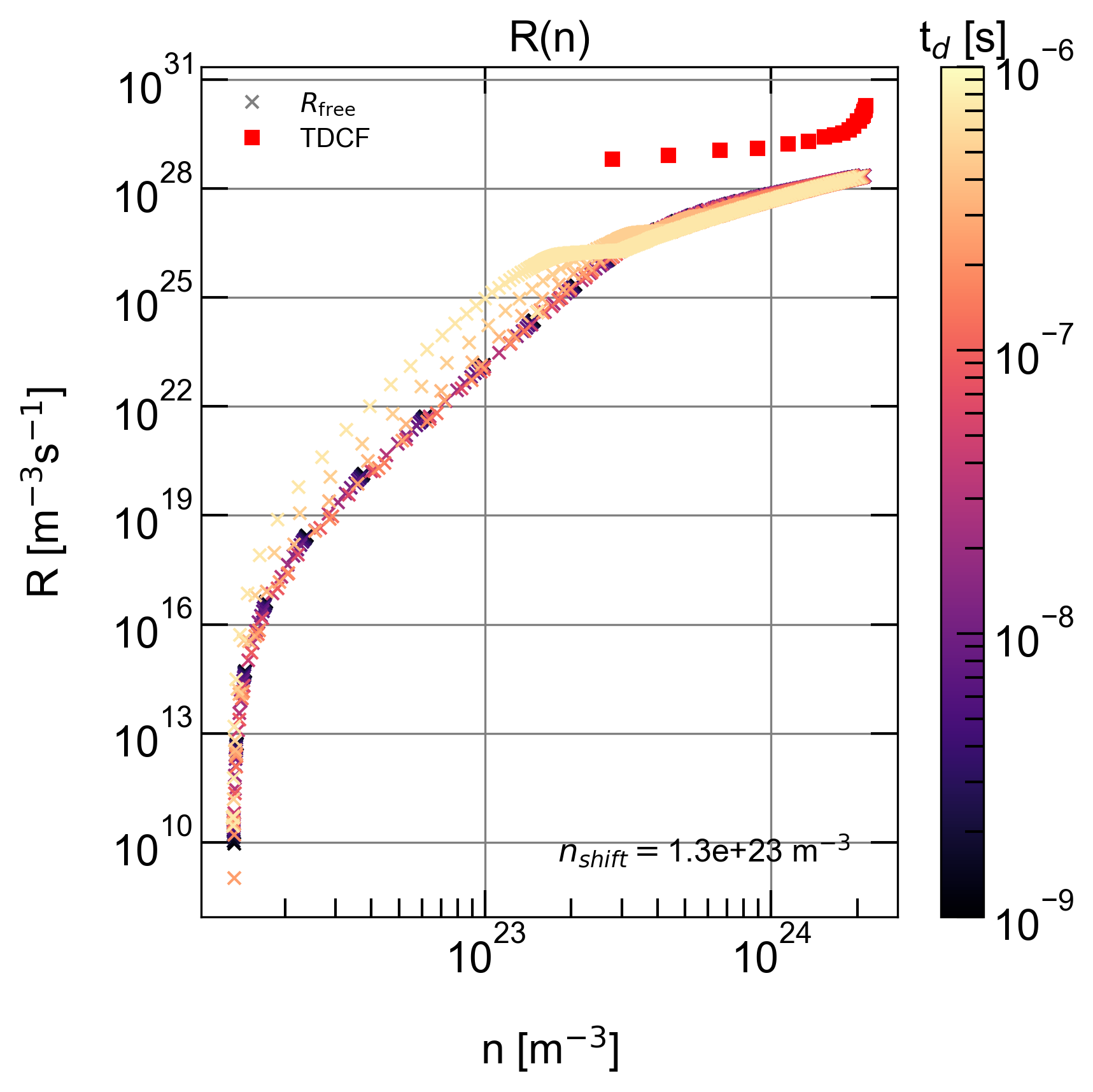}
        \caption{Comparison of TDCF-data with R(n): R is free-to-free recombination $R_{\mathrm{free}}$, no good agreement can be seen.}
        \label{fig:tdcf_r(n)_extratct_bulk_paper}
    \end{subfigure}
    \caption{Recombination in TDCF-experiments with dominating charge carrier extraction.}
    \label{fig:tdcf_extract_paper}
\end{figure}

The device in Figure 6 a) was simulated without trap states and a small bi-molecular recombination constant. The exact parameters can be found in \cref{tab:tdcf_surf_paper}. \Cref{fig:tdcf_extract_paper} d) shows that indeed the TDCF data can not be matched to the bi-molecular free to free carrier recombination rates. This is because $k_2$ is very low. Therefore charge extraction dominates the device.

\begin{table}[H]
    \caption[Simulation parameters for TDCF-experiment \textbf{Case 1}: dominating charge carrier extraction]{\label{tab:tdcf_surf_paper}Simulation parameters for TDCF-experiment \textbf{Case 1}: dominating charge carrier extraction}
    \centering
    \begin{tabular}{ll}
	    \hline\hline
	    Parameter & Value\\
	     & \\\hline
		  Active layer & PM6:Y6 $d = 10^{-7}\unit{m}$\\ \hline
		  Effective bandgap & $E_\mathrm{g} = 1.25\unit{eV}$\\ \hline
		  Cell area & $A = 0.5\cdot 0.5\unit{mm^2}$\\ \hline
		  Input mobility & $\mu_\mathrm{e} = \mu_\mathrm{h} = 10^{-7}\unit{\frac{m^2}{Vs}}$\\ \hline
		Effective charge carrier density & $n_\mathrm{eff} =5\cdot 10^{25}\unit{m^{-3}}$\\ \hline
		  Free-to-free recombination rate & $k_2 = 10^{-20}\unit{m^{3}s^{-1}}$\\ \hline
		  Laser intensity & $I = 0.1 ~\text{to}~ 5.0 \unit{\frac{\mu J}{cm^2}}$\\\hline
		  Pre-bias & $\vpre = 0.63\unit{V}$\\ \hline
		  Collection-bias & $\vcol = -5.0\unit{V}$\\ \hline
		  Delay time & $\td = 1\unit{ns}~\text{to}~10^{-6}\unit{s}$\\
		 \hline \hline
	\end{tabular}
\end{table}

\subsubsection{Recombination rates for a device with trap states}

The device in Figure 5 b) was simulated with trap states using the parameters from \cref{tab:tdcf_exptrap_paper}. An exponential density of trap states was assumed.

\begin{table}[H]
    \caption[Simulation parameters for TDCF-experiment \textbf{Case 2}: exponential distribution of trap states]{\label{tab:tdcf_exptrap_paper}Simulation parameters for TDCF-experiment \textbf{Case 2}: exponential distribution of trap states}
    \centering
    \begin{tabular}{ll}
	    \hline\hline
	    Parameter & Value\\
	     & \\\hline
		  Active layer & PM6:Y6 $d = 10^{-7}\unit{m}$\\ \hline
		  Effective bandgap & $E_\mathrm{g} = 1.25\unit{eV}$\\ \hline
		  Cell area & $A = 0.5\cdot 0.5\unit{mm^2}$\\ \hline
		  Input mobility & $\mu_\mathrm{e} = \mu_\mathrm{h} = 10^{-7}\unit{\frac{m^2}{Vs}}$\\ \hline
		  Effective charge carrier density & $n_\mathrm{eff} =5\cdot 10^{25}\unit{m^{-3}}$\\ \hline
		  Trap density & $n_\mathrm{trap} = 10^{25}\unit{m^{-3}}$\\ \hline
		  Electron, hole tail slope & $E_\mathrm{U} = 0.1\unit{eV}$\\ \hline
		  Capture cross section & $\sigma_\mathrm{trap} = 10^{-21}\unit{m^{-2}}$\\ \hline
		  Trap recombination cross section & $\sigma_\mathrm{rec} = 10^{-21}\unit{m^{-2}}$\\ \hline
		  Laser intensity & $I = 0.1 ~\text{to}~ 5.0 \unit{\frac{\mu J}{cm^2}}$\\\hline
		  Pre-bias & $\vpre = 0.63\unit{V}$\\ \hline
		  Collection-bias & $\vcol = -5.0\unit{V}$\\ \hline
		  Delay time & $\td = 1\unit{ns}~\text{to}~10^{-6}\unit{s}$\\
		 \hline \hline
	\end{tabular}
\end{table}

\end{document}